\documentclass[twocolumn]{aastex631}

\usepackage{amsmath}

\shorttitle{Peculiar Hotspot Offset of CoRoT-2b II}
\shortauthors{Kesseli et al.}
\begin{document}

\title{Unraveling the Mystery of the Peculiar and Young Hot Jupiter CoRoT-2b II: Phase Resolved Emission Spectroscopy with VLT/CRIRES+ and Gemini-S/IGRINS}

\correspondingauthor{Aurora Y. Kesseli}
\email{aurorak@ipac.caltech.edu}
\author[0000-0002-3239-5989]{Aurora Y. Kesseli}
\affiliation{IPAC, Mail Code 100-22, Caltech, 1200 E. California Blvd., Pasadena, CA 91125, USA}

\author[0000-0001-9552-3709]{Siddharth Gandhi}
\affiliation{Department of Physics, University of Warwick, Coventry CV4 7AL, UK}
\affiliation{Centre for Exoplanets and Habitability, University of Warwick, Gibbet Hill Road, Coventry CV4 7AL, UK}

\author[0000-0003-4987-6591]{Lisa Dang}
\affiliation{Waterloo Centre for Astrophysics and Department of Physics and Astronomy, University of Waterloo; Waterloo, Ontario, Canada N2L 3G1}
\affiliation{Département de physique and Institut Trottier de recherche sur les exoplanètes, Université de Montréal, C.P. 6128, Succ. Centre-ville, Montréal, H3C 3J7, Québec, Canada}

\author[0000-0001-8477-5265]{Alexis Lavail}
\affiliation{Institut de Recherche en Astrophysique et Planétologie (IRAP), 14 avenue Edouard Belin, 31400 Toulouse, France}

\author[0000-0002-0516-7956]{Alejandro S\'anchez-L\'opez}
\affiliation{Instituto de Astrofísica de Andalucía (IAA-CSIC), Gta. de la Astronomía, s/n, Genil, E-18008 Granada, Spain}

\author[0009-0002-5701-6276]{Ying Shu}
\affiliation{Waterloo Centre for Astrophysics and Department of Physics and Astronomy, University of Waterloo; Waterloo, Ontario, Canada N2L 3G1}
\affiliation{Département de physique and Institut Trottier de recherche sur les exoplanètes, Université de Montréal, C.P. 6128, Succ. Centre-ville, Montréal, H3C 3J7, Québec, Canada}

\author[0000-0003-3963-9672]{Emily Rauscher}
\affiliation{Department of Astronomy, University of Michigan, 1085 S. University, Ann Arbor, MI 48109}

\author[0000-0002-6980-052X]{Hayley Beltz}
\affiliation{Department of Physics and Astronomy, University of Kansas, Lawrence, KS 66049, USA}

\author[0000-0002-1199-9759]{Romain Allart}
\affiliation{Trottier Institute for Research on Exoplanets and D\'epartement de Physique, Université de Montréal, 1375 Avenue Thérèse-Lavoie-Roux, Montréal, QC, H2V 0B3, Canada}

\author[0000-0002-8573-805X]{Stefan Pelletier}
\affiliation{Observatoire astronomique de l’Universit\'e de Gen\`eve, 51 chemin Pegasi 1290 Versoix, Switzerland}

\author[0000-0002-2513-4465]{Vatsal Panwar}
\affiliation{School of Physics \& Astronomy, University of Birmingham,
Edgbaston, Birmingham B15 2TT, UK}



\begin{abstract}

Hot Jupiters are expected to be tidally locked and synchronously rotating due to their short orbital periods. These conditions create large day-night temperature contrasts and are thought to drive eastward super-rotating jets.
Indeed, the majority of hot Jupiters are observed to have the hottest region of the planet either at the substellar point or offset in the eastern direction. However, the full phase curve of CoRoT-2b, observed with the Spitzer Space Telescope, exhibits robust evidence of a western hotspot offset. To determine the origin of this peculiar hotspot offset, we present phase-resolved high-resolution observations of CoRoT-2b from the CRIRES+ spectrograph on the Very Large Telescope (VLT) and the IGRINS spectrograph on Gemini South, covering both pre- and post-eclipse phases (0.34--0.63). We detect the signal from the planet (S/N$>$4) in both pre- and post-eclipse phases separately, and therefore perform separate cross-correlation and retrieval analyses at the two epochs. The phase-resolved retrievals show highly consistent abundances and C/Os, but prefer a hotter and more isothermal temperature-pressure profile at post-eclipse phases, consistent with the phase curve observations that indicated a western hotpsot offset. By testing multiple hypotheses invoked to drive a western hotspot offset, we find the most likely explanation to be sub-synchronous planetary rotation. We measure the planet's rotational broadening to be $2.24\substack{+0.81\\-0.77}$ km s$^{-1}$, whereas the expectation from tidally locked rotation is $4.37\pm0.13$ km s$^{-1}$ (2.6-$\sigma$ discrepant). Other observations, such as high precision phase curves or eclipse mapping, would help to further confirm the western hotspot offset and sub-synchronous rotation.

\end{abstract}

\keywords{Classical Novae (251) --- Ultraviolet astronomy(1736) --- History of astronomy(1868) --- Interdisciplinary astronomy(804)}


\section{Introduction} \label{sec:intro}

Hot Jupiters are excellent laboratories for studying atmospheric dynamics due to their large scale heights and favorable host star flux contrasts. To date, most of the observational constraints on hot Jupiter atmospheric circulation have come from population studies of phase curves \citep[e.g., ][]{Zhang2018, Keating2019, Bell2021, Dang2025}. These studies have revealed that on average these highly irradiated planets have hot daysides and significantly cooler nightsides, and that the hottest point of the planet is located either at the substellar point or is shifted slightly to the east of the substellar point \citep{Bell2021}. These observational characteristics were predicted by global circulation models (GCMs) and can be explained by a tidally locked planet with a super-rotating equatorial jet in the direction of the planet's rotation that moves hot material east of the substellar point in the direction of rotation \citep{Showman2002, Heng2011, Rauscher2012GCM, Mayne2014}. 

Recently, high-spectral resolution observations, taken either in transmission or emission, have proven to be a powerful and complementary method for constraining atmospheric dynamics in hot Jupiters. The high spectral resolution of the instruments allow net velocity information from the planet to be resolved, and therefore wind speeds to be directly measured for the first time. In transmission, \citet{Snellen2010} used CRIRES to detect the atmosphere of HD 209458~b for the first time at high spectral resolution and noticed that the signal from the planet was blueshifted by $\sim$2 km s$^{-1}$ compared to expectations, leading the team to hypothesize that a global day-to-night wind existed on HD 209458~b. Since then net blueshifts have been measured during transmission on many hot Jupiters, and day-to-night winds are thus thought to be ubiquitous in this class of exoplanet \citep[e.g., ][]{Alonso2019a, Casasayas2019, Nugroho2020}. More recently, a time-varying radial velocity signal in planetary Fe I absorption was detected in transmission spectra of WASP-76~b \citep{Ehrenreich2020, Kesseli2021, Kesseli2022, Pelletier2023}. Numerous groups have used this first-of-its-kind detection to put constraints on atmospheric drag parameters in state-of-the-art GCMs \citep{Wardenier2021, savel2022, Beltz2022a}. Analogous time-varying signals have been measured in other ultra-hot Jupiters during transmission \citep{Simonnin2024, Wardenier2024, Lenhart2025, Langeveld2025} and can be used to understand the diversity in circulation patterns in ultra-hot Jupiters. Finally, high resolution transmission observations were used for the first time to probe atmospheric dynamics as a function of altitude by utilizing the fact that different absorption lines have different intrinsic opacities and are thus formed at different heights in the atmosphere \citep{Kesseli2024, Seidel2025}. 

Emission spectroscopy at high resolution has been used to measure or confirm hotspot offsets (or lack of hotspot offsets) by retrieving the time variation of the scaling parameter that modifies line contrast as a function of orbital phase \citep{Pino2022, vansluijs2023}. Recently, \citet{Zhang2026} mapped a half phase curve of ultra-hot Jupiter KELT-9b using KPF and measured precise radial velocities as a function of phase, allowing for tight constraints on wind speeds and leading to the measurement of a supersonic wind with speeds up to 11.7 km s$^{-1}$. Finally, signatures of atmospheric dynamics and/or chemical heterogeneities have also been suggested as the cause for offsets in observed velocities between different species in emission datasets \citep[e.g.,][]{Brogi2023, Smith2024}. Together, high-resolution transmission and emission spectroscopy are revealing the 3D nature of hot Jupiters and their atmospheric dynamics, which are then being interpreted using state-of-the-art GCMs. 

With these previous works in mind, we aim to use the power of high-resolution spectroscopy to understand the peculiar hot Jupiter CoRoT-2~b. CoRoT-2~b was discovered by \citet{Alonso2008} through the detection of transits in light curves from the CoRoT space telescope \citep{Baglin2006}, and from its discovery it has shown unusual characteristics (see Table \ref{t:corot2} for planet and stellar parameters). \citet{Alonso2008} confirmed the planetary nature of the transits with follow-up the radial velocity measurements, which allowed for precise constrains on both the mass and radius of the planet, but noted that the planet was anomalously large for its measured mass. Furthermore, secondary eclipse observations of CoRoT-2~b at a range of wavelengths have been difficult to explain using typical equilibrium models \citep[e.g.,][]{Alonso2009, Gillon2010, Snellen2010_corot2, Deming2011}, while the HST WFC3 eclipse spectrum was observed to be unusually flat \citep{wilkins2014}. Finally, a full-orbit 4.5 $\mu$m-$Spitzer$ phase curve showed a western hotspot offset \citet{Dang2018}, whereas the majority of Hot Jupiters show either no hotspot offset or an eastern hotspot offset \citep[offset in the direction of tidally locked rotation:][]{Bell2021, Dang2025}. In more recent reanalyses of $Spitzer$ phase curves, CoRoT-2~b has been the only planet to show a significant western hotspot offset that is robust against different reductions techniques \citep{Bell2021, Dang2025}. \citet{Dang2018} suggested that the western hotspot offset could be explained by western winds caused by either subsynchronous rotation of the planet \citep{Rauscher2014, Beltz2021} or magnetic effects \cite{Rogers2017, Hindle2019, Hindle2021}, and/or partial cloud coverage \citep{Roman2021, Teinturier2024}.

In the first paper in the series, \citet{Shu2025} presented high-resolution NIR observations of CoRoT-2~b at pre-secondary eclipse phases taken with IGRINS on Gemini-S. Contrary to the flat and featureless HST WFC3 eclipse spectrum, the IGRINS data showed detections of H$_2$O and CO, which represent the first detection of any molecule or atom in the atmosphere of CoRoT-2~b to date. \citet{Shu2025} subsequently perform retrievals on the IGRINS data alone, and the IGRINS data in combination with previous HST and $Spitzer$ eclipse data to constrain the atmospheric C/O and temperature-pressure profile of the planet. 

In this paper, we add additional CRIRES+ high-resolution $K$-band spectra, taken mainly at post-eclipse phases, in order to explore the phase-resolved emission spectrum. Our goals in this paper are to: (1) attempt to confirm the western hotspot offset using a different technique and dataset, and (2) distinguish between the proposed mechanisms that could create a western hotspot offset. To accomplish these goals, we perform separate cross-correlation analyses and retrievals to measure the varying pressure-temperature (PT) profile and wind parameters (i.e. wind speed, broadening) as a function of phase. The paper is laid out as follows: we present the data used in this study and describe the observation and reduction strategy in Section \ref{s:obs}. In Section \ref{s:m} we describe the analysis methods, including the cross correlation analysis (\ref{s:m_ccf}) and the atmospheric retrievals (\ref{s:m_rets}). In Section \ref{s:results} we present the findings of both the cross correlation and retrieval analyses. Finally, in Section \ref{s:disc} we discuss how they support the existence of CoRoT-2~b's western hotspot offset and the different hypotheses for the cause of the offset. Finally, we summarize our conclusions in Section \ref{s:conc}.

\begin{center}
\begin{table}[h!]
\centering
\caption{CoRoT-2 system parameters} 
\begin{tabular}{l l l}
\hline
Parameter & Value & Citation\\ 
\hline
\textbf{Star} & \\
\hline
Spectral Type & G7 & \\
$M_*$ ($M_{Sun}$) & $0.97\pm0.06$ & B17 \\
$R_*$ ($R_{Sun}$) & $0.902\pm0.018$ & B17 \\
$T_\mathrm{eff}$ (star; K) & $5625\pm120$ & B17 \\
$v_{sys}$ (km s$^{-1}$) & $23.245\pm0.01$ & A08\\
$v \sin i$ (km s$^{-1}$) & $11.85\pm0.5$ & B17 \\
$K_*$ (m s$^{-1}$) & $568\substack{+23 \\-22}$ & B17 \\
Age & $<$500 Myr & G10 \\
\hline
\textbf{Planet} &  & \\
\hline
$M_{p}$ ($M_{J}$) & $3.30\substack{+0.19 \\-0.18}$ & B17 \\
$R_{p}$ ($R_{J}$) & $1.466\substack{+0.042\\-0.044}$ &  B17 \\
$T_{eq}$ (K) & 1535$\pm$21 & O19 \\
$P_{orb}$ (d) & $1.74299709\pm0.00000007$ & A24 \\
$T_0$ (d) & $2454580.90647\pm0.00007$ & A24\\
$a$ (au) & 0.0281$\substack{+0.00057\\-0.00058}$ & B17 \\
$i$ (deg) & 88.08$\substack{+0.18\\-0.16}$ & B17 \\
$e$ & 0.0143$\substack{+0.0077\\-0.0076}$ & G10 \\
$K_p$ (km s$^{-1}$) & $175.3\pm5=3.5$ & Calculated\\
\hline
\end{tabular}
\footnotesize Note: B17 = \citet{Bonomo2017}, A08 = \citet{Alonso2008}, O19 = \citet{Ozturk2019}, A24 = \citet{Adams2024}, G10 = \citet{Gillon2010}
\label{t:corot2}
\end{table}
\end{center}

\section{Observations}
\label{s:obs}

\begin{figure}
\begin{center}
\includegraphics[width=\linewidth]{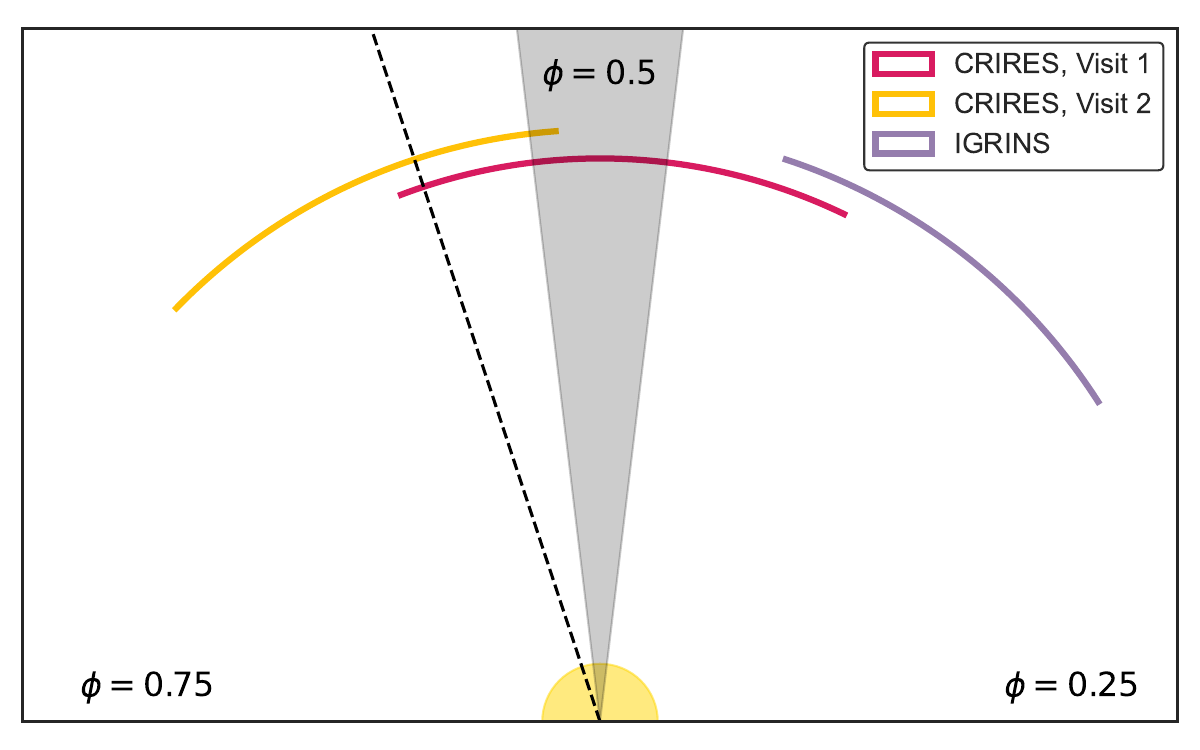}
\caption{\small
Phase coverage of the observing epochs. With the 3 epochs, we fully cover phases of 0.34 through 0.63. The two CRIRES+ epochs are newly presented in this paper, while the IGRINS epoch was first presented in \citep{Shu2025}. The gray region represents the time when CoRoT-2b is in eclipse, while the black dashed line shows the phase where CoRoT-2b has it's maximum brightness based on the observed $Spitzer$ phase curves. Our observations cover the time of peak expected brightness with about 3 hours of baseline continuing afterwards. }
\label{f:phases}
\end{center}
\end{figure}

We used observations of CoRoT-2b taken from two different NIR high resolution spectrograph to explore the peculiar western hotspot offset of the planet. In this paper, we primarily focus on the dataset from the recently upgraded Cryogenic high-resolution infrared echelle spectrograph \citep[VLT/CRIRES+; ][]{Dorn2016, Dorn2023}, but we also make use of a second dataset from the Immersion GRating
INfrared Spectrometer \citep[IGRINS:][]{Park2014, Mace2018} on Gemini-S (see Figure \ref{f:phases}). The two instruments are complimentary in nature, as CRIRES+ has higher spectral resolution,  while IGRINS has significantly wider wavelength coverage. The high spectral resolution is important for constraining the radial and rotational velocities, while the wider wavelength coverage is important for constraining chemical abundances and the pressure-temperature profile. Atmospheric retrieval results, including constraints on the C/O ratio, derived using the IGRINS dataset were presented in the first paper in this series \citep{Shu2025}. 

\subsection{CRIRES+}
\label{s:obs-crires}

\begin{figure}
\begin{center}
\includegraphics[width=\linewidth]{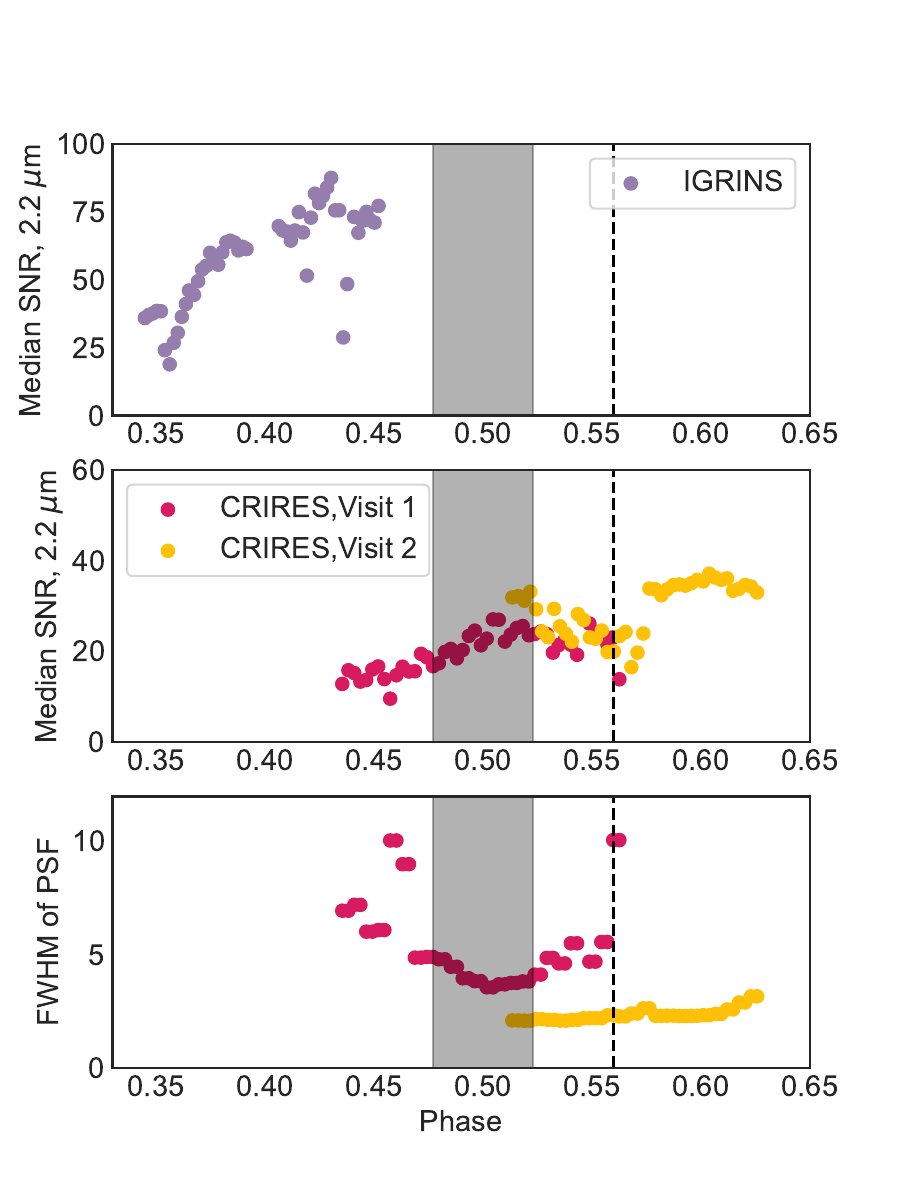}
\caption{\small 
\textbf{Top:} The median SNR of each exposure as a function of phase in the IGRINS observation epochs. We report the median SNR over all pixels in single order (order 8 in the K-band), which covers the region containing the CO bandheads, around 2.2 $\mu$m. The semi-transparent gray region shows the time when the planet was behind the star during secondary eclipse, while the black dashed line shows the time when the planet is expected to display its peak brightness from phase curve observations.
\textbf{Middle:} The same plot as above, but now showing the median SNR of the two CRIRES+ observation epochs. Again, we use a single order (order 15) with as similar wavelength coverage to the IGRINS order as possible. We see that Visit 2 had on average double the SNR per exposure as Visit 1. \textbf{Bottom:} The same plot as above, but now showing the FWHM of the PSF on the slit detector in pixels. We note that the FWHM is stable to $\pm5\%$ across the orders, and so while this plot is specifically created by fitting order 15, the plot would not change significantly for different orders. The FWHM of the PSF is used to estimate the spectral resolution when the PSF is narrower than the slit width of $\sim$3.5 pixels. The FWHMs measured during Visit 2 are consistently $\sim$2 pixels throughout the night, which corresponds to a spectral resolution of 120,000. }
\label{f:CRIRES_obs}
\end{center}
\end{figure}

We observed CoRoT-2 at two epochs using CRIRES+ as part of the observing program ID:111.24WF (PI: Kesseli). CoRoT-2 was observed for $\sim5.3$ hours on 2023-06-12, resulting in 46 spectra (Visit 1), and $\sim4.7$ hours on 2023-08-21, resulting in 42 spectra (Visit 2). Visit 1 covered phases centered around the secondary eclipse, with significant out-of-eclipse time on both sides ($0.43-0.56$), while Visit 2 covered post-eclipse phases ($0.51-0.63$; see Figure \ref{f:phases}). For both epochs, the observations were taken in an ABBA nodding pattern with exposure times of 400s. We used the narrow 0.2" slit in combination with adaptive optics to achieve the highest spectral resolution available from the slit (R$\sim$100,000). We chose to use the K2166 wavelength setting, which covers 1.92 to 2.47 microns, with gaps in the wavelength coverage, to target the strong H$_2$O absorption and CO bandheads. The instrument and setting combination has been used in previous studies to successfully detect both CO and H$_2$O in emission in similar hot Jupiters \citep[e.g., ][]{Holmberg2022, Lesjak2023, Pelletier2024}. 

The data were reduced using \texttt{cr2res}, the standard ESO pipeline for CRIRES+\footnote{\url{https://www.eso.org/sci/software/pipelines/cr2res/cr2res-pipe-recipes.html}}. First, we reduced the raw calibration files using the standard calibration cascade steps described in the DRS user manual. The raw calibrations used consist of dark frames, wavelength calibration frames (Fabry-Perot etalon and Uranium-Neon lamp), and flat-fields taken as part of the daily calibration routine. For the flat-fields, instead of the daily calibrations, we retrieved instead high-SNR deep flat fields for the K2166 setting which are acquired by ESO every few months. The calibration steps return: (a) a trace-wave file containing the geometry of the spectral orders on the science detectors and a 2-dimensional wavelength solution, (b) a normalized flat field characterizing the pixel-to-pixel flux sensitivity, and (c) a bad pixel mask.

We reduced the raw science spectra with the \texttt{cr2res$\_$obs$\_$nodding} recipe. The reduction pipeline returns a 1D wavelength calibrated and background subtracted spectrum for the A and B nodding position. The pipeline also calculates the signal-to-noise ratio (SNR) of each order and the full width at half maximum (FWHM) of the point spread function (PSF) on the detector in the slit direction. In the case when the seeing and the AO system performance are excellent, the FWHM of the PSF can be smaller than the slit width. This can cause an increased wavelength resolution (so-called super-resolution), and a shift in the wavelength solution between the A and B spectra. In that case, the spectral resolution is computed from the FWHM of the PSF, and the wavelength solution of the A and B spectra are refined using \texttt{molecfit} \citep{Smette2015} to fit the telluric lines present in the spectra. We do not use the \texttt{molecfit} telluric correction for the analysis and solely use it to refine the wavelength solution. Telluric correction is instead done using a PCA method, as is typical for the deep telluric features in the NIR and is discussed in detail in Section \ref{s:m_ccf}. We plot the median SNR and FWHM of the PSF for each exposure as a function of phase in Figure \ref{f:CRIRES_obs}. As can be seen in Figure \ref{f:CRIRES_obs}, the FWHM is stable and smaller than the slit width of 3.5 pixels throughout Visit 2, leading to a consistent super-resolution of R$\sim$120,000 throughout the night. Alternatively, in Visit 1, the FWHM is consistently above the slit width and so the resolution is set by the slit and remains at R$\sim$100,000. Due to both the higher SNR and the smaller and less variable FWHM of the PSF, the data quality during Visit 2 was significantly better than during Visit 1.

\subsection{IGRINS}

We also included in our analysis a single epoch of IGRINS data, which was observed as part of program GS-2023A-Q-111 (PI: Dang) and was first presented in \citet{Shu2025}. IGRINS simultaneously covers the H and K-bands ($1.45-2.45 \mu$m) in a single exposure at a spectral resolution of $\sim$45,000. The IGRINS observations spanned $\sim$4.5 hours on 2023-05-02 and resulted in 51 spectra over the pre-eclipse phases of $0.34-0.45$. The data were reduced using the IGRINS Pipeline Package \citep{Lee2017, Mace2018}. The package performs dark subtraction, flat fielding, wavelength calibration, and subtracts sky emission using the ABBA nodding pattern. The final pipeline product is a calibrated 1D spectrum, which still contains telluric absorption. For comparison, the median SNR for a single IGRINS order, also centered on the CO bandheads, as a function of phase is shown in a separate panel in Figure \ref{f:CRIRES_obs}. We do not plot the IGRINS and CRIRES+ data in the same panel, as the differing resolutions and instruments do not allow for an easy one-to-one comparison. More information and diagnostic plots showing the IGRINS data quality can be seen in \citet{Shu2025}.

\section{Methods}
\label{s:m}

\subsection{Cross Correlation}
\label{s:m_ccf}

\begin{figure*}[t!]
\begin{center}
\includegraphics[width=\linewidth]{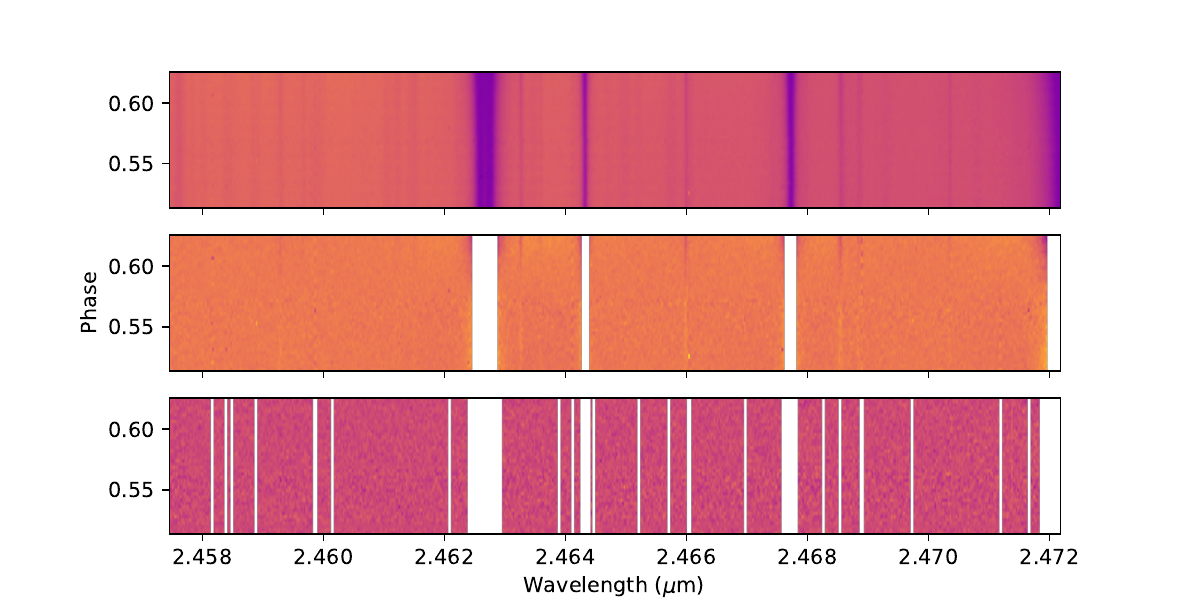}
\caption{\small 
Time series spectra for a single CRIRES+ order, showing an example of the cleaning steps that were undertaken and the final spectra. The top panel shows the pipeline reduced spectra after a normalization, sigma clipping and application of a high-pass filter. The spectra in the middle panel show the spectra after the deepest tellurics were masked and the average spectrum in time was removed. The main features that can be seen at this point are residuals from the shallower tellurics. The bottom panel shows our final cleaned spectra, after applying the \texttt{SYSREM} algorithm and masking out any columns that had large standard deviations. }
\label{f:cleaning}
\end{center}
\end{figure*}

Both the IGRINS and CRIRES+ data were cleaned and prepared for cross correlation using the same steps. For both cases the cleaning steps and cross correlation were done on an order-by-order basis, and then the cross correlation functions were combined for all of the orders. The CRIRES+ spectra consist of seven orders, with each order falling on three detectors and therefore 21 different wavelength segments. We do not attempt to combine the segments of the orders together with the CRIRES+ data and instead treat these 21 wavelength segments separately throughout our analysis. For each order segment, we first normalized the spectra to remove any variations in flux levels due to changing observing conditions over the night. We then performed a 3-$\sigma$ clipping to remove any artifacts from cosmic rays or bad pixels, and interpolated over any values that were more than 3-$\sigma$ away from pixels in the other spectra in the same pixel channel. Next, we performed filtering on the spectra in order to remove any broadband differences due to the changing blaze function over time. We followed \citet{Kesseli2022} and \citet{Merritt2020}, and divided each spectrum by the average of the observed time series. We then applied a Gaussian filter with a width of 200 pixels to the resulting average-removed spectra. The original spectra were then divided by their corresponding filtered spectra in order to remove any low-frequency noise, while preserving the higher-frequency signal from the exoplanet and the shape of the spectrum. The spectra for one example spectral order at this step are shown in the top panel of Figure \ref{f:cleaning}. 

Next, we performed steps to mitigate telluric contamination. For all the data, we began by simply removing the deepest telluric lines by masking any columns where the flux was below 40\% of the maximum value in that order. As in many previous studies \citep[e.g.,][]{Birkby2017, Nugroho2017, Cabot2019, Kesseli2020}, we iteratively applied the \texttt{SYSREM} algorithm \citep{Tamuz2005, Mazeh2007} to remove systemic trends that were quasi-stationary in time, while preserving the rapidly Doppler shifting signal of the exoplanet. \texttt{SYSREM} cannot be applied blind, and removing too many \texttt{SYSREM} iterations has been shown to also remove the exoplanet's signal \citep{Birkby2017}. However, fine-tuning the number of \texttt{SYSREM} iterations too much has also been shown to create spurious signals or artificially inflate the resulting signal-to-noise ratio (SNR) \citep[see ][]{Cabot2019}. We therefore chose to use the same number of  \texttt{SYSREM} iterations on all orders. We chose to use 5 \texttt{SYSREM} iterations, as this was the fewest number of iterations required to remove signs of quasi-stationary trends near the telluric lines via visual inspection of the corrected spectra and resulting cross correlation functions. We also tested using between 3 and 10 \texttt{SYSREM} iterations and found that the resulting signal or signal properties (radial velocity, strength, etc.) did not change drastically, as expected given the results of previous studies \citep[e.g., ][]{Alonso2019a, SanchezLopez2019, Pelletier2024}. As a final step, we masked any spectral columns with high variance in the \texttt{SYSREM}-corrected spectra, which resulted in masking between $2-8\%$ of the channels, depending on the order. An example of the final cleaned spectra that are fed into our cross-correlation algorithm can be seen in the bottom panel of Figure \ref{f:cleaning}.

\begin{figure}
\begin{center}
\includegraphics[width=\linewidth]{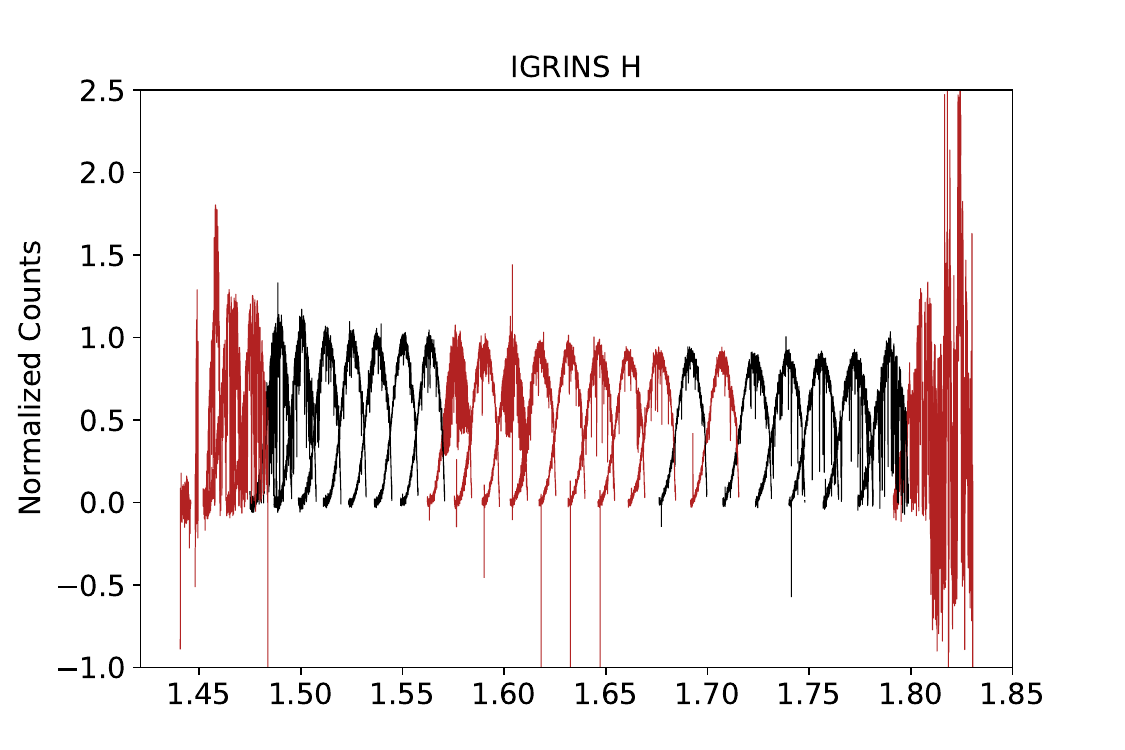}
\includegraphics[width=\linewidth]{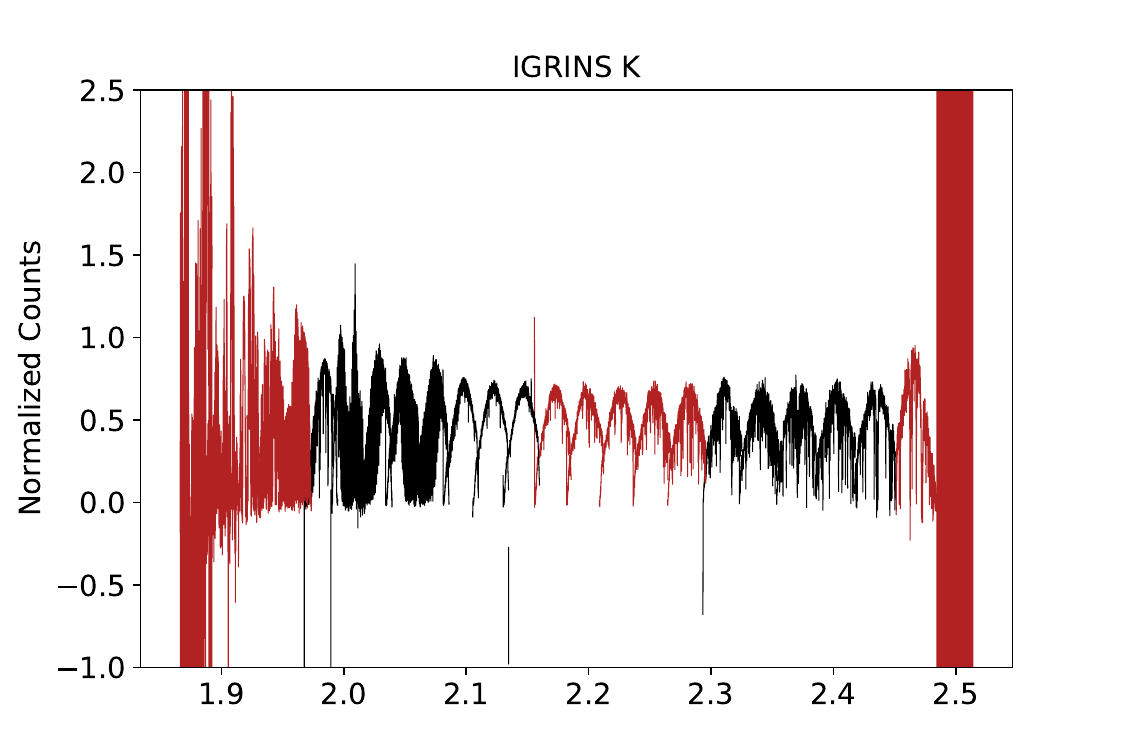}
\includegraphics[width=\linewidth]{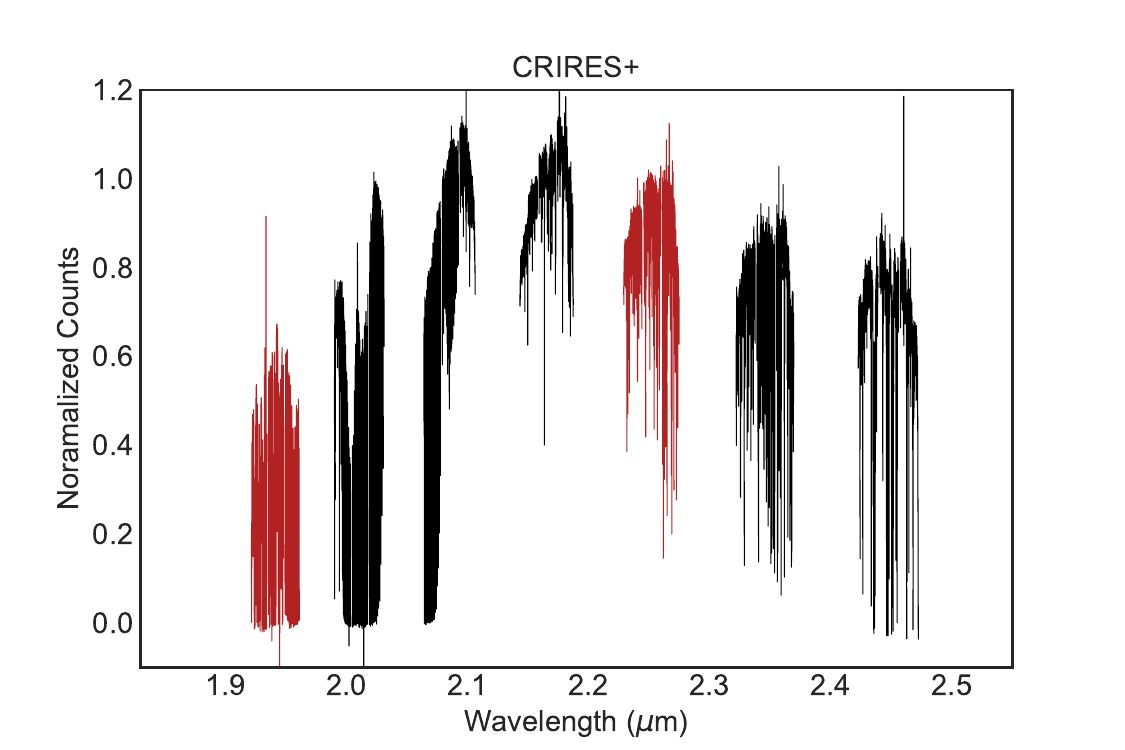}
\caption{\small 
 Representative spectra of CoRoT-2 taken with IGRINS in the H-band (top), K-band (middle), and with CRIRES+ using the K2166 setting (bottom). These plots show all orders of a single time series spectrum produced by the data reduction pipeline steps before any spectral cleaning has been done. The orders plotted in black are the orders which are used in the final combined 2D cross-correlation grids, while the orders in red are thrown out as they did not produce a large signal in our injection and recovery tests. The lack of signal is due to either low SNR of the orders (i.e. orders at the edge of the detector), telluric contamination, or orders where there is not significant opacity from CO or H$_2$O (i.e. around 1.65 or 2.2 microns). We also found significant CH$_4$ contamination in the Visit 2 CRIRES+ spectra in the region around 2.25 microns, and so we opted to discard these data as well. }
\label{f:data_orders}
\end{center}
\end{figure}

We generated model spectra of the atmosphere of CoRoT-2b in emission to use as cross correlation templates using petitRADTRANS \citep{Molliere2019} in the high spectral resolution mode. The model spectra output by petitRADTRANS are given as the planet flux (F$_\mathrm{p}$), and so we divided by a blackbody at the star's effective temperature to obtain F$_\mathrm{p}$/F$_\mathrm{s}$. Since the first paper in the series (\citet{Shu2025}) measured the strongest signals from H$_2$O and CO, we made templates with each of these two molecules separately, and one that contained both H$_2$O and CO. We used the latest ExoMol opacities for H$_2$O \citep[Pokazatel:][]{Polyansky2018}, which have been shown to fit exoplanet spectra well \citep{Gandhi2020}. petitRADTRANS also requires other inputs, such as the pressure-temperature (PT) profile, the planet's radius and surface gravity, abundances, and the mean molecular weight of the atmosphere. We used the retrieved values from \citet{Shu2025} for the PT profile (see discussion of the Guillot PT in Section \ref{s:m_rets}) and abundances, and
the values from the literature listed in Table \ref{t:corot2}. The resulting model spectra were then broadened to the two instruments' spectral resolutions (R=100,000 and R=120,000 for CRIRES+ nights 1 and 2, respectively, and R=45,000 for IGRINS). 

We cross correlated the models with the cleaned time-series spectra following \citet{Hoeijmakers2020} and \citet{Kesseli2021} to perform the cross correlation analysis, using the following equation: 

\begin{equation} 
\label{e:1}
c(\nu, t) = \sum_{i=0}^{M} - x_i(t) T_i(\nu). 
\end{equation}

\noindent where $c(\nu, t)$ is the resulting 2D cross-correlation grid at each radial velocity ($\nu$) and time of observation ($t$). $x_i(t)$ are all of the individual spectra, while $T_i(\nu)$ is the model, shifted from $\nu = -250$ to $+250$ km s$^{-1}$ in 0.5 km s$^{-1}$ steps. The models are normalized such that $\sum_{i=0}^{M} T_i(\nu) = 1$, and a negative sign is included so that excess absorption from the exoplanet will appear as a positive residual, as is standard for cross correlation. 

\begin{figure}[h!]
\begin{center}
\includegraphics[width=\linewidth]{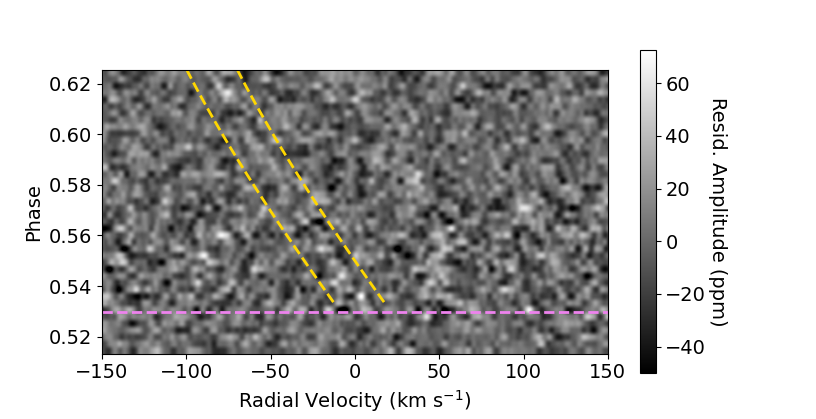}
\caption{\small 
2D cross correlation grid for the CRIRES+ Visit 2 dataset in Earth's rest frame. Each row is a resulting cross correlation function between the model spectra of CoRoT-2~b that contained CO and H$_2$O and a cleaned time-series spectrum. Larger residual amplitudes indicate positive correlation between the model and the data. The horizontal violet dashed line indicates the end of secondary eclipse, while the yellow dashed lines show the planet's expected velocity. A faint region of increased residual amplitudes between the gold dashed lines shows that the planet is detected. }
\label{f:2D_CCF_CRIRES}
\end{center}
\end{figure}

This step was repeated separately for every order and the orders were then combined by simply averaging the 2D cross-correlation grids ($c(\nu, t)$) together. We did not perform any complex weighting of the orders, except for multiplying by 0 (for orders not included) and 1 (for the orders included). To determine which orders to include in the combined cross-correlation grids, we injected the planetary model that contained H$_2$O and CO into the spectra at five times the expected strength and at a systemic velocity of $+100$ km s$^{-1}$ (far from the true planet signal) and performed all the same cleaning and cross-correlation steps described above. If the injection test resulted in an SNR $>$ 5 in that order, we included it in the final averaged 2D cross-correlation grid. Figure \ref{f:data_orders} shows which orders resulted in injected signal strengths with an SNR $>$ 5, and were therefore included in the final analysis for each spectrograph. An example showing the final order-combined 2D cross correlation grid for the CRIRES+ Visit 2 data is shown in Figure \ref{f:2D_CCF_CRIRES}. This dataset showed the highest overall SNR, and the trace of the planet can be faintly made out even in the 2D grid without co-adding all the exposures together. 

\subsection{Retrievals}
\label{s:m_rets}

We performed retrievals on the CRIRES+ Visit 2 (post-eclipse) spectra so that we could compare the retrieved parameters after secondary eclipse to the retrieved parameters presented in \citet{Shu2025} that correspond to phases before secondary eclipse. 
%The goals of the retrievals were to: (1) add further evidence to support the existence of the western hotspot offset, (2) compare abundances and cloud opacities at the two epochs to determine if the data supports clouds on the eastern side of the planet, and (3) derive a value of the rotational broadening of the planet's signal. 

We used the HyDRA-H framework \citep{Gandhi2018, Gandhi2019_hyrdra} for our retrievals of the CRIRES+ data, which uses MultiNest \citep{Feroz2009, Buchner2014} to perform nested sampling using 300 live points and a changing sampling efficiency to accurately explore the peaks of the posterior. HyDRA-H has been successfully used for high resolution retrievals in a range of studies \citep[e.g., ][]{Gandhi2020,Line2021, Gandhi2023}. HyDRA-H has also been extensively validated against retrievals that use petitRADTRANS for the spectral model generation, and shown to produce highly consistent results \citep{Ahrer2025}. The retrieval framework assumes free parameters for atmospheric chemistry (each species is allowed to vary in accordance to the value preferred by the data), but keeps the abundance of each species constant in altitude. We adopt line lists from ExoMol \citep{Tennyson2024} for H$_2$O \citep{Polyansky2018}, CH$_4$ \citep{Yurchenko2024} and HCN \citep{Harris2006, Barber2014}, and HITEMP \citep{Rothman2010} for CO and CO$_2$. We also use the H- opacity due to bound-free and free-free interactions from \citet{Bell1987} and \citet{John1988}. Each line is broadened according to the pressure (Lorentzian) and temperature (Gaussian), resulting in a Voigt profile \citep[e.g.][]{Gandhi2020_cs}. In addition, we include a term for the Doppler broadening due to the planet's rotation, which is left as a free parameter.

We use two separate temperature profile parametrizations for our retrievals. We firstly adopt a modified profile from \citet{Guillot2010}, with five free parameters, as is implemented in petitRADTRANS \citep{Molliere2019}. This is the same temperature profile used for the retrieval performed in paper 1 \citet{Shu2025}. We also adopt the \citet{Madhusudhan2009} profile, with six free parameters with three distinct regions in the atmosphere. Both of these profiles are flexible enough to constrain non-inverted, isothermal and inverted temperature profiles from our data. The use of both verified our results were consistent and not dependent on the choice of parametrization used. The prior ranges for each parameter are shown in Table~\ref{t:ret}.

In addition to gaseous absorption, we also include a contribution from absorption due to any cloud species present. This is parametrized as a gray opacity $\kappa$, with the vertical extent of the absorption given by 
\begin{align}
        \kappa_\mathrm{cl}(P, \lambda) = 
    \begin{cases}
      \kappa_{\mathrm{cl}} \left(\dfrac{P}{P_\mathrm{cl}}\right)^{\alpha_\mathrm{cl}} & P\leq P_\mathrm{cl}, \\
      0 & P> P_\mathrm{cl}. \\
    \end{cases}  
\end{align}
Here, $P_\mathrm{cl}$ refers to the cloud deck pressure and $\alpha_\mathrm{cl}$ is the cloud opacity power law, both of which are free parameters in the retrieval. 

Finally, we also added free parameters for $\delta K_p$ and $\delta v_{sys}$, which are the deviation from the planet’s known orbital velocity and a deviation from the system's known velocity (Table \ref{t:corot2}). Table \ref{t:ret} lists the priors for all the parameters in the retrievals.  
For every model iteration, we generated a spectrum at a spectral resolution of 500,000 in the wavelength range of CRIRES+ data and then convolved down to the instrumental resolution for Visit 2 (120,000). %The model spectrum was further convolved with a rotational broadening kernel, where the value for the rotational broadening was fit for as a free parameter. 

\begin{table*}[ht]
\caption{Priors and retrieved posteriors for the different retrieval runs. } 
\begin{tabular}{l l | l | l l l}
\hline
& Parameter & Prior & Guillot PT, All & Guillot PT, Detected & M\&S$^\dagger$ PT, All\\ 
\hline
& log$_{10}$H$_2$O & U(-15.0, -1.0) & $-4.28\substack{+1.97\\-0.74}$ & $-4.96\substack{+0.55\\-0.40}$ & $-4.16\substack{+1.33\\-0.78}$ \\
& log$_{10}$CO & U(-15.0, -1.0) & $-4.14\substack{+1.68\\-1.16}$ & $-4.75\substack{+0.91\\-0.84}$ & $-3.97\substack{+1.47\\-1.0}$ \\
Abundances & log$_{10}$CO$_2$ & U(-15.0, -1.0) & $<-3.08$$^\ddagger$ & ... & $<-3.61$ \\
& log$_{10}$CH$_4$ & U(-15.0, -1.0) & $<-1.95$ & ... & $<-2.51$ \\
& log$_{10}$HCN & U(-15.0, -1.0) & $<-3.78$ & ... & $<-4.15$ \\
& log$_{10}$H$^-$ & U(-15.0, -1.0) & $<-4.73$ & ... & $<-5.67$ \\
\hline
 &log$_{10}$P$_{cl}$ (bar) & U(-4.0, 2.0) & $-0.59\substack{+1.68\\-2.00}$ & $-0.33\substack{+1.61\\-2.42}$ & $-0.64\substack{+1.68\\-2.12}$ \\
 Clouds &log$_{10} \kappa$ (cm$^2$g$^{-1}$) & U(-10.0, 10.0) & $-1.83\substack{+5.71\\-5.47}$ & $-3.04\substack{+5.88\\-4.38}$ & $-2.46\substack{+6.5\\-5.1}$\\
 & log$_{10}\alpha_{cl}$& U(0.0, 20.0) &$10.7\substack{+5.4\\-6.1}$ & $11.0\substack{+5.7\\-6.6}$ & $11.0\substack{+5.9\\-6.5}$ \\
\hline
&T$_{eq}$ (K) & U(300.0, 4500.0) & $2577\substack{+273\\-284}$ & $2403\substack{+206\\-377}$ & ...\\
&$\alpha$ & U(-1.0, 1.0) & $0.18\substack{+0.29\\-0.51}$ & $-0.10\substack{+0.51\\-0.57}$ & ... \\
Guillot PT&log$_{10}\delta$ & U(-6.0, 2.0) & $-5.11\substack{+0.64\\-0.60}$& $-5.26\substack{+0.72\\-0.53}$& ... \\
&log$_{10}\gamma$ & U(-2.0, 2.0) & $-0.39\substack{+0.34\\-0.35}$& $-0.44\substack{+0.23\\-0.36}$ & ... \\
&log$_{10}P_{trans}$ & U(-5.0, 3.0)& $-3.36\substack{+1.5\\-1.1}$& $-2.81\substack{+2.91\\-1.34}$& ... \\
\hline
&T$_{set}$ (K) & U(300.0, 4500.0) & ... & ... & $2593\substack{+145\\-128}$\\
&$\alpha$1 & U(0.0, 1.0)& ... & ... & $0.48\substack{+0.20\\-0.16}$\\
M\&S PT&$\alpha$2 &U(0.0, 1.0) & ... & ... & $0.49\substack{+0.28\\-0.18}$\\
&P1 & U(-6.0, 2.0) & ... & ... & $-1.99\substack{+1.7\\-1.9}$\\
&P2 & U(-6.0, 2.0) & ... & ... & $-3.7\substack{+2.1\\-1.5}$\\
&P3 & U(-2.0, 2.0) & ... & ... & $0.60\substack{+0.94\\-1.26}$\\
\hline
 &$\delta$ K$_p$ (km s$^{-1}$) & U(-20.0, 20.0) & $-0.51\substack{+2.64\\-3.21}$& $-0.21\substack{+2.78\\-3.48}$ & $-0.80\substack{+2.80\\-3.86}$ \\
Velocities &$\delta$ V$_{sys}$ (km s$^{-1}$) & U(-20.0, 20.0) & $0.81\substack{+1.58\\-1.87}$& $1.04\substack{+1.54\\-1.98}$& $0.70\substack{+1.62\\-2.18}$\\
 &$v \sin i$ (km s$^{-1}$)& U(1.0, 15.0) & $2.24\substack{+0.81\\-0.77}$& $2.07\substack{+0.95\\-0.68}$ & $2.16\substack{+0.968\\-0.76}$\\
\hline
\end{tabular}
\\
\footnotesize $^\dagger$ M\&S = \citet{Madhusudhan2009} \\
\footnotesize $^\ddagger$ all limits are 2-$\sigma$ upper limits
\label{t:ret}
\end{table*}

As input into the retrievals, we used the CRIRES+ Visit 2 data that had already been run through the data reduction pipeline, but had not had any spectral cleaning steps applied. We do not use the final cleaned spectra from the cross correlation analysis (i.e., masked and detrended with \texttt{SYSREM}), as we will perform cleaning steps within the retrieval so that the same steps are performed on both the model and data to not bias the retrieval. We used the same orders/wavelength segments as in the cross correlation analysis (all black wavelength segments in the bottom panel of Figure \ref{f:data_orders}), totaling 15 wavelength segments. As in the cross-correlation analysis we perform the cleaning on each wavelength segment separately and discard the bluest wavelength regions due to lower SNRs and telluric H$_2$O contamination, and the three wavelength segments around 2.2$\mu$m due to residual telluric CH$_4$ contamination. Although we found it necessary to discard these telluric-contaminated wavelength segments, it may mean that our retrievals are less sensitive to other species that abosorb in these areas, such as CH$_4$ and HCN. We also discard the five exposures that were taken during secondary eclipse/egress. 

To detect any signal from the planet, some cleaning of the data was required. We performed the cleaning steps within the retrieval so that every step that was performed on the data was also performed on our generated model spectra. We followed the PCA-based cleaning method presented in \citet{Panwar2024}, which is optimized for short computation times and therefore more suitable for use within a retrieval than our cleaning steps performed within the CCF analysis. We refer the reader to \citet{Panwar2024} for details, but provide a general overview of the steps here. We first 'standardized' the time-series spectra by subtracting the mean and dividing by the standard deviation in each channel so that large differences in flux between the
continuum and deep telluric lines would not
bias the PCA towards either. The PCA is then performed on the 'standardized' data cube using \texttt{numpy}'s single value decomposition to find the set of N$_{\mathrm{PCA}}$ eigenvectors that can describe the data (where N$_{\mathrm{PCA}}$ is the number of PCA components/eigenvectors). We chose to use 5 PCA components, as this number of PCA components was tested in our cross correlation analysis. 
Since the PCA was performed on the standardized data, an additional step of multi-linear regression to the original data using the PCA eigenvectors is required to compute the best fit model to the stellar and telluric systematics in the data. When doing multi-linear regression, we also append a 'bias vector' of ones to the set of PCA components to account for the mean-subtraction during standardization. Finally, we divide the original data by this multi-linear regression model fit and are left with cleaned time series spectra in each order. This method was extensively validated in \citet{Panwar2024, Panwar2025}, but we also performed our own validation of the cleaning steps and saved the final cleaned spectra from the retrieval pipeline and performed a cross correlation analysis on them and found that we could recover the signal with a similar, but slightly lower SNR as was achieved with our cross correlation analysis. Because the cleaning within the cross correlation analysis provided a slightly higher SNR, we do still opt to use two different cleaning methods for our two different analyses, and also note that this shows that the signal is robust towards different methods of cleaning the data. 

To ensure that the impact of PCA cleaning on the underlying planetary signal in the data are also emulated in the forward model used for cross-correlation analysis, we take the commonly adopted approach of 'model-reprocessing' by following the same steps described in detail in \citet{Panwar2024, Panwar2025}. In brief, for every pair of K$_p$ and V$_{sys}$, we compute the total planetary velocity at each orbital phase in the data, Doppler shift the forward model and inject it additively into the data at each time step. We then repeat the PCA cleaning step on the model injected data using the same eigenvectors as those for cleaning the data, to yield cleaned model injected data. We next subtract the cleaned data from the cleaned model injected data, to yield the reprocessed time series of forward models. This reprocessed model time-series encodes the impact of PCA cleaning on the planetary signal in the data, and can be used to compute more accurate cross-correlation derived log-likelihood during retrievals.

\section{Results}
\label{s:results}

\subsection{Cross Correlation}

\begin{figure}[h!]
\begin{center}
\includegraphics[width=0.8\linewidth]{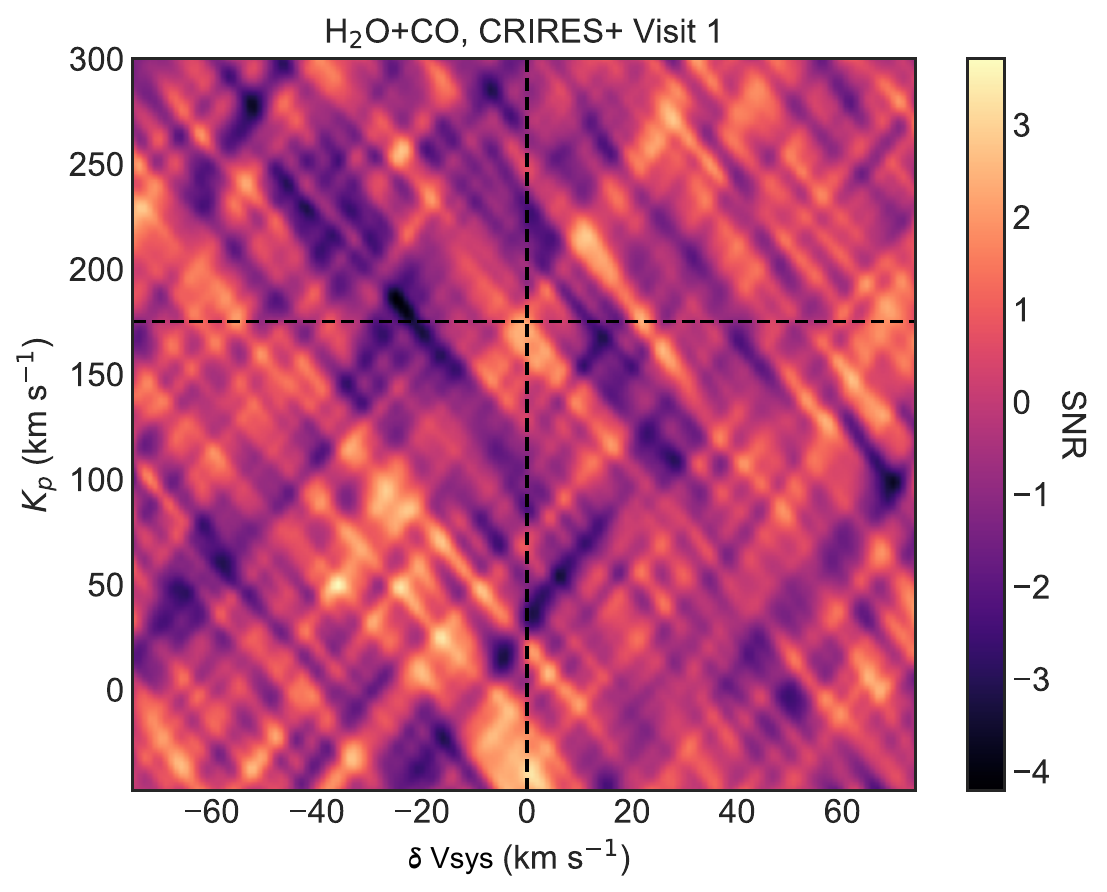}
\includegraphics[width=0.8\linewidth]{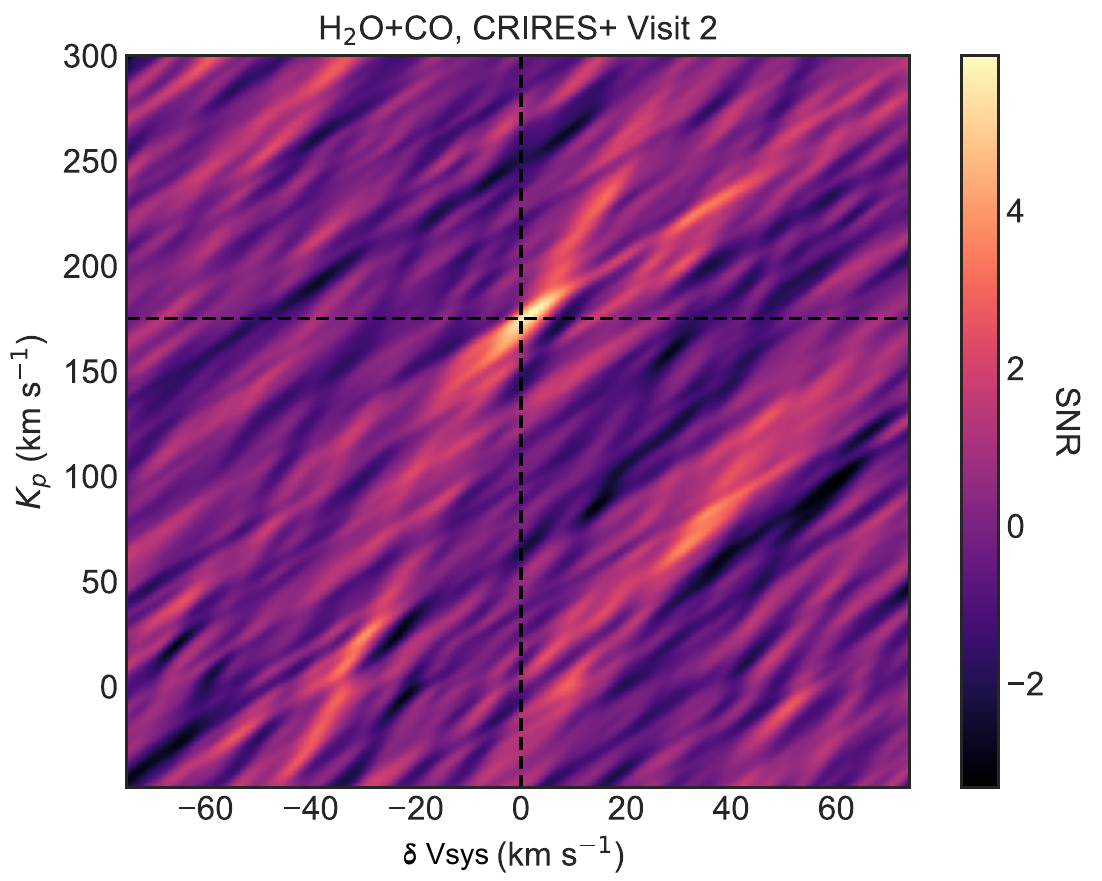}
\includegraphics[width=0.8\linewidth]{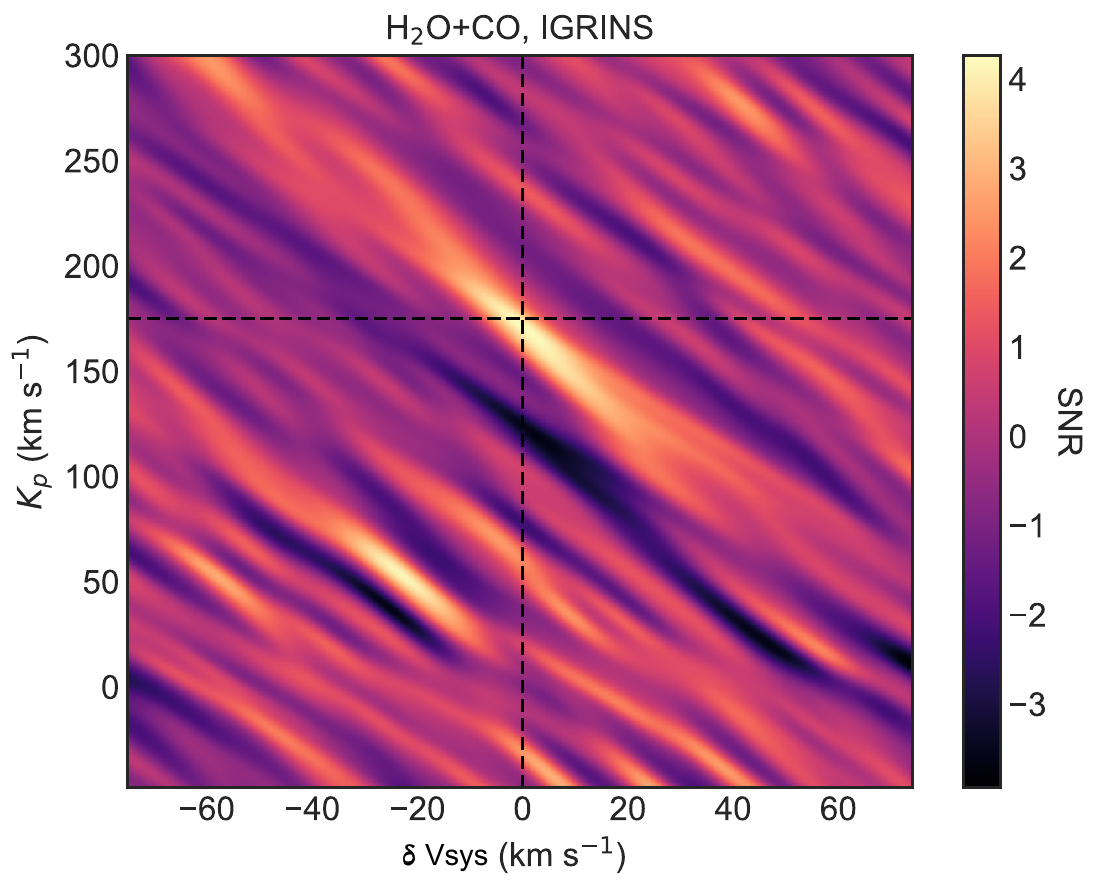}
\caption{\small 
K$_p$ vs. $\delta v_{sys}$ diagrams for each observation epoch. The CRIRES+ Visit 2 epoch and the IGRINS epoch both exhibit the strongest signal at the planet's expected location (dashed black lines) and exhibits SNRs$>$4 at that location, and so we can confidently say that we have detected the planet at these epochs. In the CRIRES+ Visit 1 epoch we do see a signal at the expected position, but it is only at an SNR$\sim$3 and is not the strongest signal we observe, and so we are not able to confidently detect the planet. We therefore proceed in the rest of the analysis only using CRIRES+ Visit 2 and the single IGRINS epoch. }
\label{f:KpVsys}
\end{center}
\end{figure}

%A common method to determine the strength of the detection and ensure that there are no spurious signals is to make K$_p$ vs. $\delta v_{sys}$ diagrams. 
To assess the strength of the planet's signal in each separate epoch, we created K$_p$ vs. $\delta v_{sys}$ diagrams. The diagrams were created by assuming a range of different planet semi-amplitudes (K$_p$) and radial velocities ($\delta v_{sys}$) to shift the 2D cross correlation grid (see Figure \ref{f:2D_CCF_CRIRES}) to the planet's rest frame. We calculated the planet's velocity at each observation ($v_p(t)$) using the equation: 
\begin{equation}
   v_p(t) = v_{sys} + \delta v_{sys}+ v_{bary}(t) + \mathrm{K}_p \sin (2\pi \phi(t))
\label{e:vel}
\end{equation}

\noindent where $v_{sys}$ is the known systemic velocity (see Table \ref{t:corot2}), $\delta v_{sys}$ is an additional velocity shift on top of the known systemic velocity, $v_{bary}(t)$ is the barycentric correction calculated at each observation $(t)$, and $\phi(t)$ is the phase, which is also calculated at each observation $(t)$. $\phi(t)$ is calculated using the equation: 
\begin{equation} 
\phi(t) = (t - \mathrm{T_0}) / \mathrm{P}
\label{e:phi}
\end{equation}
\noindent where P is the period and T$_0$ is the time of conjunction. Once in the planet's rest frame, all of the individual CCFs are simply co-added to make a 1D CCF and the SNR is recorded by taking the value at 0 km s$^{-1}$ and dividing by the standard deviation far from the center ($>+50$ or $<-50$ km s$^{-1}$). 

If there is a signal from the planet, a maximum in the SNR should appear near the known K$_p$ (175 km s$^{-1}$, Table \ref{t:corot2}) and at a $\delta v_{sys}$ close to 0 km s$^{-1}$. A slight offset (up to $\sim$10 km s$^{-1}$) from the expected position is possible due to winds and dynamics in the planet's atmosphere, but significant signals far away from the expected position are due to noise and/or tellurics that have not been properly removed and likely signify that a detection should not be trusted. The final diagrams for each epoch are shown in Figure \ref{f:KpVsys} and demonstrate that the planet is confidently detected for CRIRES+ Visit 2 and the IGRINS visit. We do not confidently detect the planet in the CRIRES+ Visit 1 epoch, and so we do not use it for the rest of the analysis or in the retrievals. This non-detection for the CRIRES+ Visit 1 epoch is not unexpected given the lower quality of the data (see Figure \ref{f:CRIRES_obs}), and a hint of the signal at SNR$\sim$3 at the expected planet's velocity is also consistent with this picture. 

\subsubsection{Radial Velocity vs. Phase}
\label{s:r_rvphase}

With the planet confidently detected at pre-eclipse phases in the IGRINS dataset and post-eclipse phases with the CRIRES+ Visit 2 dataset, we move on to using the CCFs to measure the velocity offset from the expected planetary velocity ($\delta v_{sys}$) as a function of phase. Significant $\delta v_{sys}$ values have been constrained in numerous exoplanet atmospheres and have been used to constrain planetary winds and circulation patterns \citep[e.g.,][]{Snellen2010, Ehrenreich2020, Kesseli2021}. 
Equatorial winds, which are thought to drive hotspot offsets, can impart velocity shifts on the order 1-6 km s$^{-1}$ in emission spectra of hot Jupiters \citep{Zhang2017, Beltz2024_survey}. Measuring the wind velocity as a function of phase, instead of averaged over all phases, can be especially diagnostic because as the planet rotates and the hotspot comes into and out of view, the measured velocity offset will change. By measuring $\delta v_{sys}$ as a function of phase and comparing to the output of GCM models \citep[e.g.,][]{Zhang2017, Beltz2024_survey}, we aim to constrain the equatorial wind velocity and direction to determine if the observations are consistent with western winds driving a western hotspot offset.

To measure any excess velocity due to winds in the planet's atmosphere, we must precisely know the planet's true orbital velocity ($v_p(t)$). We can calculate $v_p(t)$ using Equations \ref{e:vel} and \ref{e:phi}, where $\delta v_{sys}$ is now kept at 0 km s$^{-1}$, as this excess $\delta v_{sys}$ is what we aim to measure. 
It is crucial to have accurate and precise known velocities ($v_{sys}$, K$_p$) and timing information (P, T$_0$) because uncertainties or inaccuracies in these parameters can propagate into $v_p(t)$, which would then manifest in inaccurate or imprecise measured planetary wind values \citep[e.g.,][]{Pai2022}. Therefore, to confidently detect and set constraints on atmospheric winds, these quantities need to be treated carefully. 

As CoRoT-2 is a very young and active star, measuring accurate transit time centers has plagued this system since its discovery \citep{Alonso2008}. The initial 80 consecutive transits from CoRoT showed timing variations on the order of 20s due to stellar activity from the young host star \citep{Alonso2008}, leading to imprecise values for T$_0$. Other subsequent studies saw evidence for a non-linear ephemeris, potentially indicating tidal decay, which requires more complicated calculations than Equation \ref{e:phi} for calculating $\phi(t)$ \citet{Ozturk2019, Ivshina2022, Wang2024}. However, in the most recently published paper, \citet{Adams2024} performed a careful reanalysis of all archival data in combination with new TESS transits and corrected errors in some previously reported reference time frames and found no evidence of deviation from a linear ephemeris. The newly derived T$_0$ and P values lead to a propagated uncertainty of less than 25 seconds for the dates observed, leading to an uncertainty on the calculated velocities of $<0.2$ km s${^-1}$.
$v_{sys}$ was reported by \citet{Alonso2008} to be 23.245 km s$^{-1}$, with an uncertainty of only 0.01 km s$^{-1}$, which causes a negligible effect on the total error budget. Finally, we calculate K$_p$ using the following equation, K$_p = 2 \pi a \sin i$/P, where $a$ is the semi-major axis, $i$ is the inclination, and P is the period. Using values from the most recently published parameter update \citep{Bonomo2017}, and propagating the uncertainty in each value, we calculate a K$_p$ of 175.3$\pm$3.5 km s$^{-1}$. Using other values from the literature we find values of K$_p$ that range from 174.5 to 177.6, in good agreement with our preferred value. By propagating all of the above uncertainties into our pipeline, we find differences in our measured velocity offsets of at most 1.5 km s$^{-1}$. 

\begin{figure}[h!]
\begin{center}
\includegraphics[width=\linewidth]{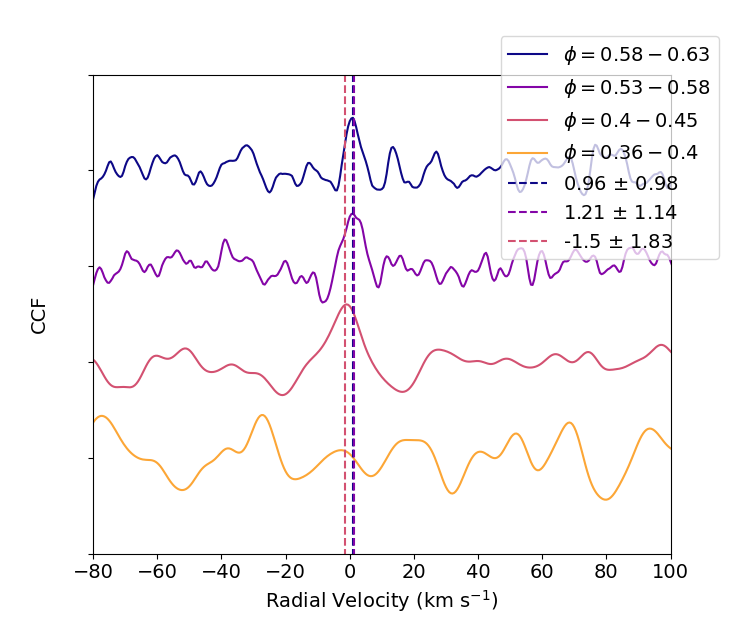}
\caption{\small 
1D co-added CCFs in the planet's rest frame at four different phase epochs. The CCFs are offset by a constant coefficient for clarity. The two bottom CCFs (orange and red lines) are both from the IGRINS visit, while the two top CCFs (purple and blue) are from CRIRES+ Visit 2. We do not measure a significant planet signal for pre-eclipse phases less than 0.4, but measure significant peaks at the other phase bins (with SNRs of 4.6, 4.0, and 4.2 for the CCFs from top to bottom). For all three phase bins our measured radial velocity offset is consistent with zero to within 1.5-$\sigma$, although we do measure a non-significant slight blueshift pre-eclipse and slight redshift post-eclipse.    }
\label{f:RVvsPhase}
\end{center}
\end{figure}

Using the calculated values of $v_p(t)$ at each exposure, we shifted each CCF in the 2D CC grid (i.e., Figure \ref{f:2D_CCF_CRIRES}) into the planet's rest frame, and co-added all the CCFs into two phase bins for the CRIRES+ post-eclipse epoch and another two phase bins for the IGRINS pre-eclipse epoch. The 1D co-added CCFs for the four phase bins are shown in Figure \ref{f:RVvsPhase}. We find that between the pre-eclipse phases of 0.36 and 0.4 we are not able to detect the signal from the planet. \citet{Shu2025} also found a lack of signal at these orbital phases. This lack of signal is consistent with expectations given the lower SNR of the spectra at these phases (see Figure \ref{f:CRIRES_obs}) in combination with the rotation of the planet's hotter dayside away from the observer. Between phases of 0.4 and 0.45 we measure a maximum in the planet's signal with an offset from the planet's rest velocity of $\delta v_{sys} = -1.5\pm1.83$ km s$^{-1}$. \citet{Shu2025} measure $\delta v_{sys} = -3.2\pm1.72$ km s$^{-1}$, which is within one sigma of our measurement. At the two post-eclipse phases, we alternatively measure a positive velocity offset of $\delta v_{sys} = +1.21\pm1.14$ km s$^{-1}$ and $+0.96\pm0.98$ km s$^{-1}$ for phase bins $0.53-0.58$ and $0.58-0.63$, respectively. We find that all of the measured $\delta v_{sys}$ values are consistent with 0 at the $\sim$1-$\sigma$ level. 

\subsubsection{Planet Rotational Broadening}
\label{s:r_broad}
Using the CRIRES+ Visit 2 1D phase-combined CCF, we measured the planet's $v\ sin(i)$ to estimate the planet's rotation and determine whether the planet may be rotating faster or slower than expectations assuming tidal locking. We used a method similar to \citet{Kesseli2018}, and opted to only use the data from CRIRES+ Visit 2, as this epoch showed a clear detection of the planet and achieved the highest spectral resolution (R$\sim$120,000), therefore allowing the best constraints on broadening. 

\begin{figure}
\begin{center}
\includegraphics[width=0.9\linewidth]{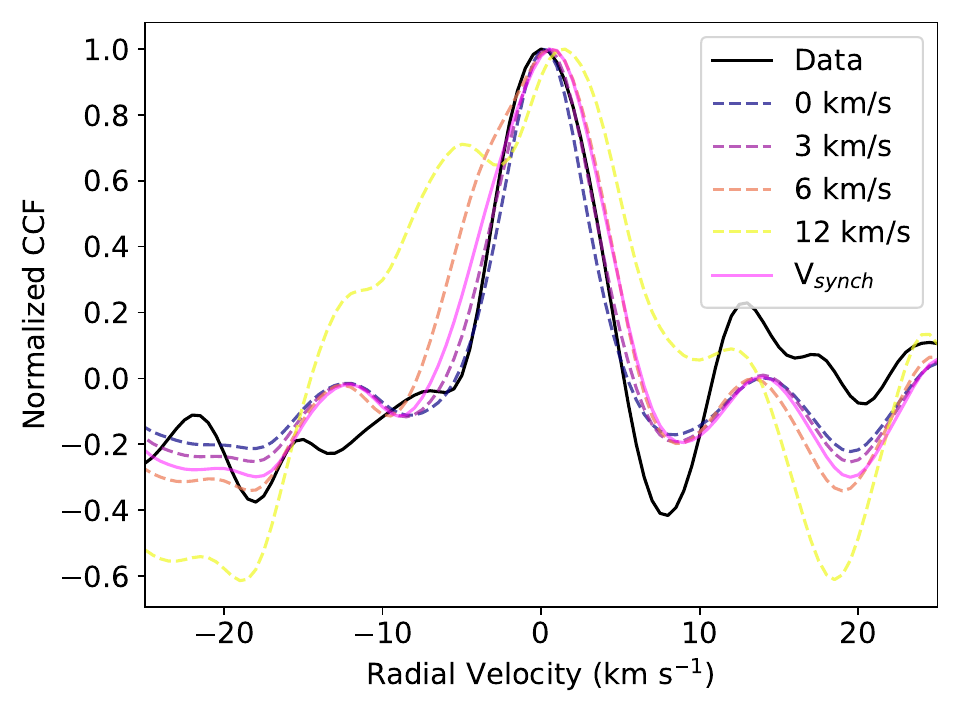}
\caption{\small 
Comparison between the 1D CCFs resulting from the cross-correlation between the artificially broadened models and the unbroadened model (dashed lines) versus the data with the same unbroadened model (black solid line). The models were broadened with a $v\ sin(i)$ kernel and correspond to $v\ sin(i)$ values from 0 to 12 km s$^{-1}$. The models and the data have all been normalized so that their peak values are at 1 so that the widths of the lines can be compared more easily. The width of the CCF made using the CRIRES+ data falls between the profiles of the 0 and 3 km s$^{-1}$ $v\ sin(i)$ models, and we determine a $v\ sin(i)$ of 2.1 km s$^{-1}$ using this method. }
\label{f:RotBroad}
\end{center}
\end{figure}

We created a model of the planetary atmosphere using petitRADTRANS and the known planetary parameters. We then artificially broadened this model using the $v\ sin(i)$ kernel from the pyAstronomy library\footnote{\url{https://pyastronomy.readthedocs.io/en/latest/pyaslDoc/aslDoc/rotBroad.html}} at steps of 2 km s$^{-1}$, starting with $v\ sin(i)$ values of 0 and ending at 12 km s$^{-1}$. We then reduced the resolution of the resulting spectrum down to the resolution of CRIRES+ in Visit 2, by convolving with a Gaussian kernel and interpolating onto the same wavelength grid as the data. The resolution of the CRIRES+ instrument is stable enough in time and across wavelength that this simple method is sufficient. We finally cross correlated each artificially broadened model with the unbroadened ($v\ sin(i) =$0 km s$^{-1}$) model and measured the FWHM of each CCF. These values calibrated a relationship between FWHM and $v\ sin(i)$, which we could then compare with the data. We created an analogous 1D CCF by cross-correlating the data with the same unbroadened model and combining all out-of-eclipse CCFs in the planet's rest frame and measured the FWHM. Our measured FWHM from the data resulted in a planetary $v\ sin(i)$ of 2.1 km s$^{-1}$. Figure \ref{f:RotBroad} shows resulting 1D CCFs from the data, as well as different broadened models.

\subsection{Retrievals}

Our atmospheric retrievals for the CRIRES+ dataset converge to tight constraints on many parameters (see Appendix \ref{A:ret}), including $\delta K_p$ and $\delta v_{sys}$ values near zero ($\delta K_p = -0.29\substack{+2.17\\-2.81}$ km s$^{-1}$, $\delta v_{sys} = 0.93\substack{+1.20\\-1.47}$ km s$^{-1}$) as expected. The retrieved values match expectations for a planet with CoRoT-2b's properties (temperatures, detected chemical species), and we find consistent physical parameters (PT profiles, abundances, velocities) when the model parameterizations are slightly altered (consistent results between the three different retrieval setups in Table \ref{t:ret}). Furthermore, the retrievals also show analogous results with the cross-correlation analysis for the velocity parameters, and both $\delta v_{sys}$ and $v\ sin(i)$ are consistent within 1-$\sigma$ between our two analyses. All of these separate checks lead us to conclude that the retrievals strongly detect the signal from the planet and that they can be used to accurately constrain planet properties. 

As the planet's rotational broadening is one of the main parameters of interest in this study, we used the retrievals to obtain another separate measurement of the rotational broadening as a secondary check of its accuracy. We find in our retrievals that the rotational broadening is well constrained by our data, and for our nominal retrieval run (Guillot PT, All) we find a value of $2.24\substack{+0.81\\-0.77}$ km s$^{-1}$. This value is consistent between the different retrieval set-ups (see Table \ref{t:ret}). We also note that the only degeneracy between the retrieved rotational broadening value and any of the other retrieved parameters is with the other velocities ($\delta$ K$_p$, $\delta$ V$_{sys}$, and that the rotational broadening goes up only as $\delta$ K$_p$ and $\delta$ V$_{sys}$ are offset farther from their known values (e.g., as $\delta$ K$_p$ goes to more negative values; see Appendix \ref{A:ret}), indicating that the errorbars are robust or even slightly overestimated for our $v\ sin(i)$. 

\begin{figure*}
\begin{center}
\includegraphics[width=0.59\linewidth]{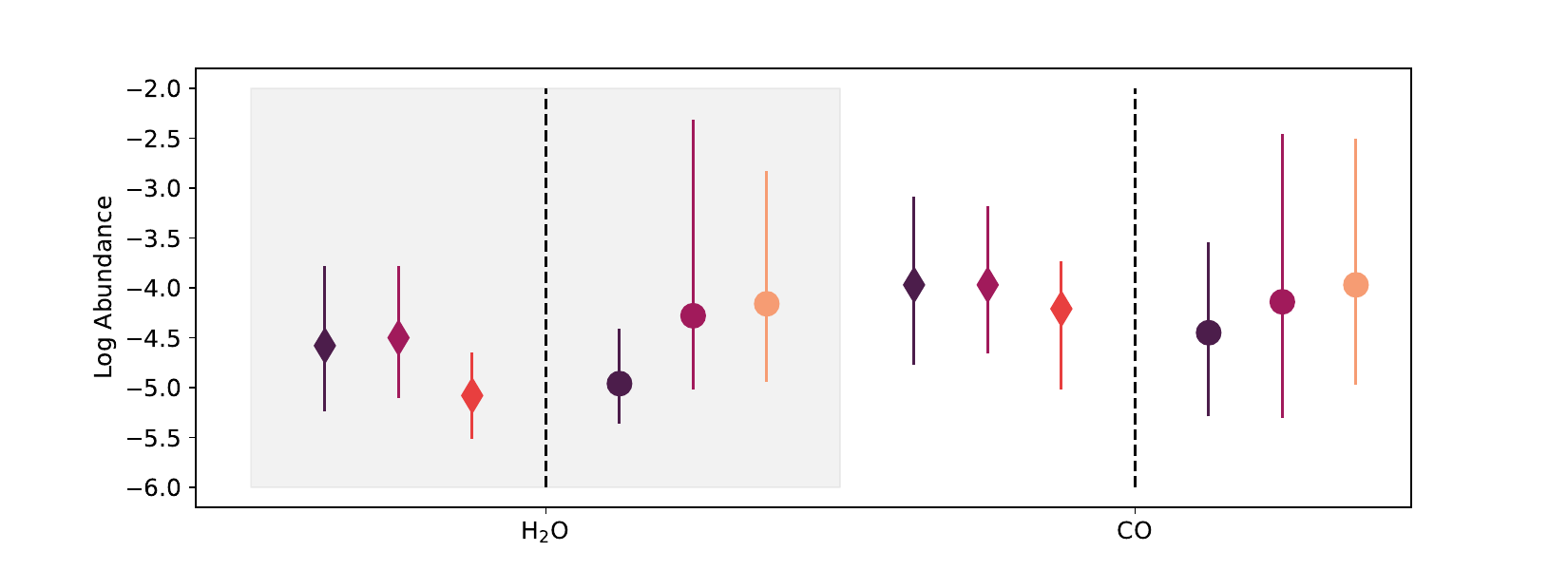}
\includegraphics[width=0.40\linewidth]{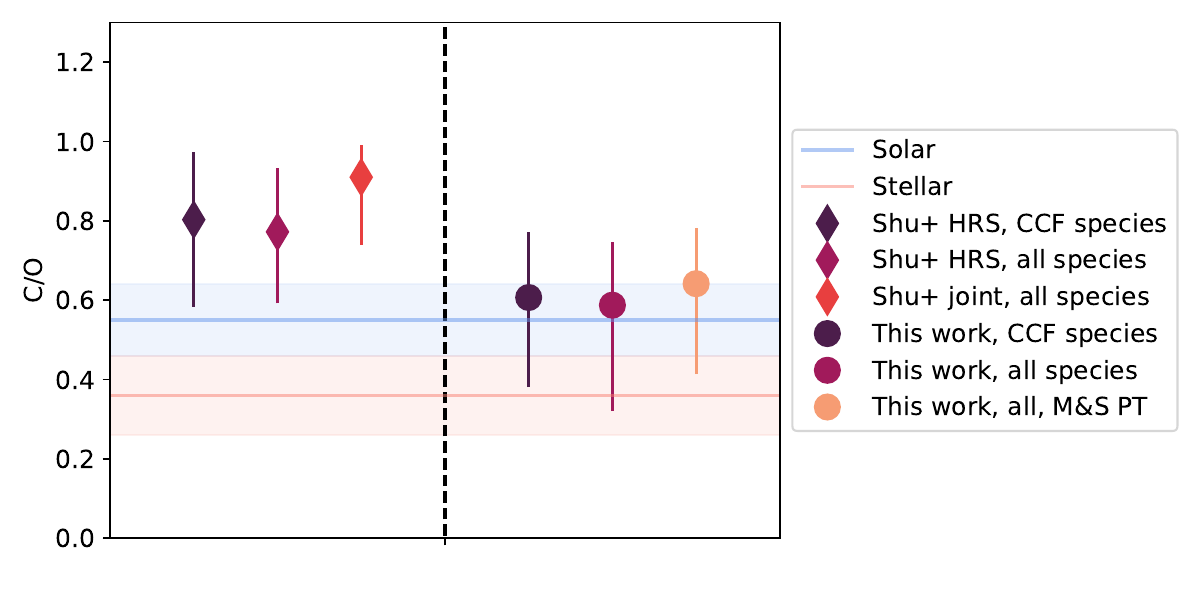}
\caption{\small 
Comparison between the retrieved abundances (left) and abundance ratios (C/O; right) from the retrieval presented in \citet{Shu2025} (diamonds) and our CRIRES+ retrievals (circles). We compare to all three retrieval set ups presented in \citet{Shu2025}, including the two using the high-resolution pre-eclipse IGRINS data only (HRS) and the joint IGRINS+HST+Spitzer retrieval (joint).} We find that while the abundances vary between the different datasets and retrieval setups, the abundance ratios are highly consistent and all point towards a solar or slight super-solar C/O. \textbf{The solar and stellar C/Os and their uncertainties are plotted as a horizontal blue line \citep{Asplund2009} and a horizontal orange line \citep{daSilva2024}, respectively.  CoRoT-2 has a measured C/O that is slightly sub-solar \citep[$0.36\pm0.11$; ][]{daSilva2024}, which makes the potentially elevated carbon abundance slightly more robust. }
\label{f:abunds}
\end{center}
\end{figure*}

Although not the focus of this paper, we also compared the chemical abundances we derive from the post-eclipse CRIRES+ retrievals performed here to the pre-eclipse IGRINS retrievals from \citet{Shu2025}. We find consistent measured abundances of CO and H$_2$O across the different instruments/retrieval pipelines and different model conditions (see Figure \ref{f:abunds}), which is very promising. The abundances result in consistently super-solar C/O ratios, although within uncertainties of the solar value \citep[$0.55\pm0.09$][]{Asplund2009}. The [C/Fe] and [O/Fe] ratios were recently measured for CoRoT-2 \citep{daSilva2024}, and results in a slightly sub-solar C/O of $0.36\pm0.11$, which may be common for young stars \citep{Takeda2017}. In Figure \ref{f:abunds} we see that both our retrievals and those in \citet{Shu2025} find slightly higher abundances of CO compared to H$_2$O, which leads to slightly super-solar/stellar C/O ratios in all cases (values between $\sim0.6-0.9$). This agreement in the abundance ratios between the two datasets and subsequent retrievals is encouraging and gives us confidence that conclusions drawn from comparing the retrievals can be trusted.

\section{Discussion}
\label{s:disc}

We used the results from both the cross-correlation analysis and the retrieval analysis to: (1) attempt to provide additional evidence for the existence of the western hotspot offset, and (2) to distinguish between the potential physical scenarios that were invoked to explain the peculiar hotspot offset. 

\subsection{Evidence for Western Hotspot Offset}

In order to test the hypothesis of a western hotspot offset, we compared the retrieved PT profile from the post-eclipse CRIRES+ dataset to the PT profile retrieved in \citet{Shu2025} for the pre-eclipse IGRINS data. Retrievals are the best way to determine the hotspot offsets from high-resolution spectroscopy because it is not straightforward to use the strength of the CCFs to deduce an atmospheric temperature. As was exemplified in \citet{vansluijs2023}, high-resolution spectroscopy is more sensitive to the temperature gradient than the actual temperature of the atmosphere, and so while the temperature is expected to increase at the hottest point on the planet, the temperature gradient becomes shallower and the strength of the measured CCF can actually decrease when the hottest region of the planet is in view. 

\begin{figure}[h!]
\begin{center}
\includegraphics[width=\linewidth]{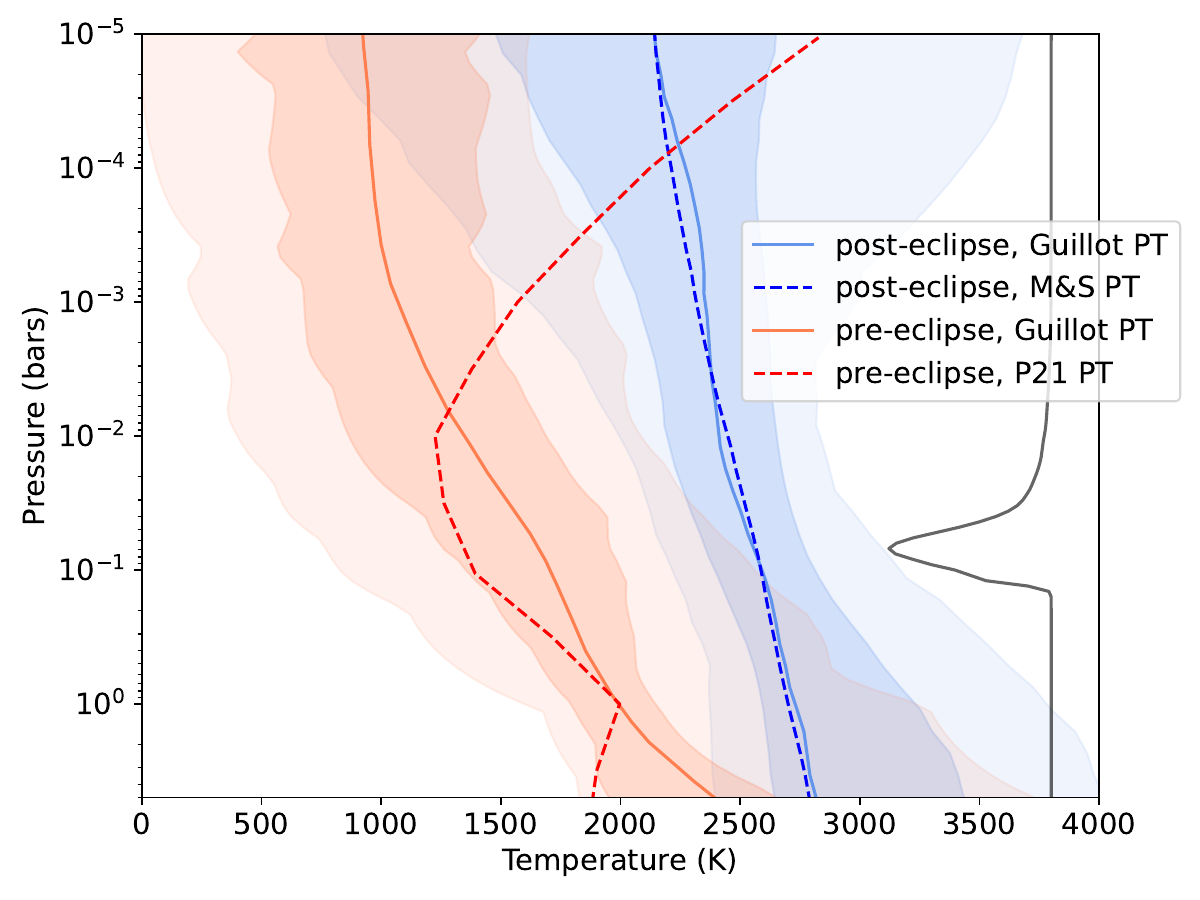}
\caption{\small 
Retrieved PT profiles using the post-eclipse CRIRES+ data (blues) and the pre-eclipse IGRINS data (reds), published in \citet{Shu2025}. The profiles plotted with the solid lines are both parameterized using the same modified Guillot formulation, specified in \citep{Molliere2019}. The shaded regions correspond to the 1-$\sigma$ and 2-$\sigma$ uncertainty regions for the Guillot profiles. The profiles with the dashed lines use a free parameterization with pressure anchor points. For the IGRINS data this profile was fit using the \citet{Pelletier2021} parameterization, and for the CRIRES+ data the \citet{Madhusudhan2009} parameterization. The black solid line shows a contribution function for the CRIRES+ Guillot PT retrieval, to demonstrate where in altitude the observations are most sensitive. The different PT parameterizations show consistent results within the region where the signal originates, demonstrating that the differences in the pre- and post-eclipse PT profiles is unlikely due to any choice in the PT profiles. We find that the post-eclipse data is better fit with a hotter PT profile that is more isothermal, whereas the pre-eclipse data prefers a cooler temperature but with a larger gradient in temperature change (a less steep PT profile).}
\label{f:PTs}
\end{center}
\end{figure}

The retrieval performed on the IGRINS data in \citet{Shu2025} and the nominal retrieval performed on the CRIRES+ data presented here use the same modified Guillot parameterization as specified in \citet{Molliere2019}, making the comparison between the two PT profiles straightforward. We do not include the retrievals from \citet{Shu2025} that included HST and Spitzer data, as those are not confined to pre-eclipse phases and are therefore not suitable for compare the pre- and post-eclipse PT profiles. We plot the different PT profiles in Figure \ref{f:PTs}, and show that we retrieve both a hotter and more isothermal PT profile in the post-eclipse data than in the pre-eclipse data at the pressure ranges probed. As the slope of the PT profile and the presence of clouds can also be degenerate, we also ran a retrieval identical to our normal model but without any clouds and find an almost identical PT profile (always within 1-$\sigma$ of the nominal model), adding to the validity of the PT profile. We also plot the retrieved PT profiles from the pre- and post-eclipse data when different PT parameterizations are used, and find that the different PT parameterizations show consistent results over the pressure ranges that are probed by the data (black contribution function) and that regardless of the PT parameterization used, the post-eclipse data show a hotter and more isothermal temperature structure. This matches expectations from GCM models, where the hottest regions on the planet are expected to have hotter temperature but more isothermal PT profiles \citep{Showman2015, vansluijs2023}. We take this as evidence that our observations are consistent with a western hotspot offset, but we note that the uncertainties in the PT profiles are large and so this dataset alone cannot provide independent confirmation of a western hotspot offset.

\subsection{Mechanism driving the hotspot offset}

\citet{Dang2018}, proposed three possible causes for the western hotspot offset: magnetic effects, sub-synchronous rotation, or clouds. In the following sections we attempt to use the results of our analyses to comment on whether the data support each hypothesis. 

\subsubsection{Magnetic Effects}

Both strong magnetic effects and sub-synchronous rotation could drive winds and/or a jet westward, opposite to the direction of rotation \citep{Rauscher2014, Beltz2022b}. If the magnetic field is driving western winds, we would expect both the winds and hotspot offset to be variable \citep[e.g.,][]{Rogers2017}, whereas if sub-synchronous rotation is driving the western winds we would expect little variability. Our data is consistent with a western hotspot offset, and so this may indicate a lack of variability, or it could simply indicate that our observations happen to occur during another epoch with a western hotpsot offset. With only one phase curve epoch and our tentative confirmation of a western hotspot offset at a second epoch, there are few conclusions we can draw from a variability standpoint.

Another avenue for assessing the magnetic scenario and constraining the magnetic field strength could be to compare the velocity offset of the cross correlation function as a function of phase (see Figure \ref{f:RVvsPhase}), as proposed in \citet{Beltz2024_survey}. Unfortunately, it is not straightforward to compare the velocity shifts as a function of phase to MHD models and recover constraints on the magnetic field strength because the MHD models used are not actually able to produce a hotspot offset reversal, as this requires a non-ideal MHD framework to support regions where the magnetic Reynolds number exceeds unity \citep{Beltz2022a, Rogers2017}. Future updates to MHD frameworks in combination with detailed modeling of the conditions of CoRoT-2~b may make this possible in the future. 

Finally, we assess whether we expect strong magnetic effects in CoRoT-2~b's atmosphere and revisit the field strength required to cause a reversal of the jet towards the western direction on CoRoT-2~b. Magnetic effects are predicted to become more pronounced in hotter planets due to increased thermal ionization \citep{RauscherMenou2013, Kennedy2025}. Specifically, \citet{Kennedy2025} show that it is not until equilibrium temperatures of 2000K and above that magnetic drag begins to disrupt the eastern jet, leading to a hotspot offset progressively closer to the substellar point. In the case of CoRoT-2~b, a reversal (not simply a damping) of the jet is required to observe a western hotspot offset. With the requirement of a field reversal in combination with an equilibrium temperature on the cooler end of where magnetic effects become relevant, \citet{Dang2018} estimated that a magnetic field of strength B$\sim$230 G would be required to cause the observed hotspot offset. While magnetic fields have not been directly measured on exoplanets, \citep{Yadav2017} estimated hot Jupiter magnetic field strengths based on scaling arguments for inflated hot Jupiters and found that magnetic fields up to $\sim$$50 - 100$G may be possible. Recent evidence for magnetic fields on hot Jupiters via phase curve observations found values on the order of $\sim$$1-7$G \citep{Kreidberg2018, Rogers2017}. Lastly, the most recent MHD GCMs have used field strengths ranging from $0-30$G, as the likely range of expected magnetic field strengths on hot Jupiters. Given weaker magnetic field strengths typically assumed for hot Jupiters, the 230G field required for CoRoT-2~b's hotspot reversal seems outside of the expected range and therefore relatively unlikely. 

Recent work has suggested that magnetic effects may drive hotspot offset variability for cooler planets through thermoresistive magnetic instabilities \citep[TRI;][]{Hardy2022, Hardy2023}. This instability, which arises from the large  differences of electrical conductivity as a function of temperature, occurs when the ohmic dissipation rate is faster than the radiative cooling timescale \citep{Menou2012}. However, \citet{Hardy2023} suggest that sustained oscillations are expected when $T_{eq} \lesssim 1200$K, which is significantly cooler than CoRoT-2b, and we therefore do not expect TRIs to significantly shape the hotspot location for CoRoT-2b.

\subsubsection{Sub-synchronous Rotation}

As with magnetism, sub-synchronous rotation has also been shown to be able to drive a western jet in the opposite direction of rotation \citep{Rauscher2014, Beltz2021}. We calculated the planet's rotational velocity assuming synchronous rotation with the following equation: $v_{synch} = 2 \pi R_p / P$. Using values from Table \ref{t:corot2} we calculated $v_{synch}$ to be 4.37$\pm0.13$ km s$^{-1}$. The retrieved rotational broadening value and the measured value from the cross correlation analysis are both smaller than $v_{synch}$. Using the value from the nominal CRIRES+ retrieval ($2.24\substack{+0.81\\-0.77}$km s$^{-1}$), we find that it is inconsistent with the calculated $v_{synch}$ value, albeit by only 2.6-$\sigma$. 

Even though the IGRINS data does not have the spectral resolution to resolve sub-synchronous rotation (R$\sim$45,000 corresponds to instrumental broadening on the order of $4-7$km s$^{-1}$), we can ensure that we do not find broadening $greater$ than the instrumental resolution as an additional consistency check. Using the same CCF method discussed in Section \ref{s:r_broad}, we measure a planet $v \sin i = 4.6$km s$^{-1}$. This value is as expected for rotational broadening at or lower than the spectral resolution of the instrument, and is therefore consistent with the CRIRES+ measurement and a low rotational broadening for the planet. 

\citet{Beltz2021} present GCM simulations of varying rotation speeds (including sub-synchronous rotation) in the canonical hot Jupiter, HD 209458~b, and find that the full broadening of the planet's absorption lines are relatively constant and larger expectations from synchronous rotation due to the combination of rotation and an eastern jet (see Figure 9 in \citealt{Beltz2021}). Only the slowest rotating models (P$\sim 1/2$P$_{locked}$) drive winds in the western direction, leading to both a western hotspot offset and broadening significantly smaller than that which is expected from synchronous rotation. Our low $v \sin i$ may then be interpreted as additional evidence for a western flow, which counteracts the eastern rotation to produce a very low $v \sin i$ and acts to drive a western hotspot offset. While the velocities presented here are only marginally inconsistent, we note that previous measurements of planetary rotational broadening in both hot and ultra-hot Jupiters have consistently measured significantly broader planetary lines than what would be expected for synchronous rotation \citep[e.g.,][]{Lesjak2023, Lesjak2025}. CoRoT-2~b is therefore an outlier in both its western hotspot offset and its narrow planetary lines, indicating that a western flow is indeed present in the atmosphere and adding credence to the sub-synchronous rotation hypothesis. 

While sub-synchronous rotation would result in a planet $v \sin i$ that is smaller than $v_{synch}$ and is consistent with expectations from GCM models, we also considered whether the small $v \sin i$ could instead be caused by the signal originating primarily from a small hotspot and not well distributed across the surface of the planet. We derive a very simple order of magnitude estimate for how small the hotspot must be assuming the retrieved $v \sin i$ and then calculating the hotspot radius via R$_\mathrm{hotspot} = v \sin i \times P/2\pi$, and find R$_\mathrm{hotspot}=0.8$R$_\mathrm{J}$ (compared to the measured 1.466R$_\mathrm{J}$ value). If the signal originated from a hotspot this small, then we would expect to see significant radial velocity variations as the hotspot rotated into and out of view. Our results from Section \ref{s:r_rvphase} do not show any significant radial velocity variations as a function of phase, and so we conclude that a hotspot is unlikely to be the cause for the low $v \sin i$.

As in the magnetic field scenario, we aim to next assess whether sub-synchronous rotation is feasible for CoRoT-2~b. It is commonly assumed that all hot-Jupiters with zero eccentricity are tidally locked based on the argument that the synchronization timescale is smaller than the circularization timescale \citet{Rasio1996}. CoRoT-2~b is known to have a small, but non-zero eccentricity \citep[$e = 0.0143\substack{+0.0077 \\-0.0076}$;][]{Gillon2010}, which may indicate ongoing tidal processes. While CoRoT-2 is known to be a young host star (age estimates range from 100 to 300 Myr range: \citealt{Schroter2011, Guillot2011}), the common paradigm is that a planet that is not yet synchronized would be spinning faster and then slow down to become tidally locked. Indeed, given the system parameters we would expect CoRoT-2~b to be tidally locked even at its young age, with tidal locking timescales on the order of a few Myr (see Appendix \ref{B: tidal_lock} for calculation details). However, even if CoRoT-2~b was in a synchronous rotation state at one point, \citet{Showman2002} argue that the feedback of angular momentum between the orbit, atmosphere, and interior can torque exoplanets out of synchronous rotation. Thermal tides \citep{Arras2009, Arras2010, Auclair2018} and magnetic torque \citep{Wazny2025} have been suggested as two processes that may drive planets away from synchronous rotation. Because synchronization of hot Jupiters is so widely assumed and there has not been strong observational evidence indicating otherwise, there has been limited work on processes to torque hot Jupiters out of synchronization, and so we conclude that while there seem to be mechanisms that can cause sub-synchronous rotation, more work needs to be done to assess the physical likelihood of these processes and encourage future theoretical studies motivated by these results. 

Finally, we explore a last scenario where the planet itself could be inclined (e.g., rotating on its side). \citet{Malsky2021} show that an inclined planet would indeed lead to a low $v \sin i$, as we have measured. An inclined spin axis may be the result of the planet being trapped in Cassini state 2 \citep[e.g.,][]{Winn2005, Fabrycky2007}. However, \citet{Levrard2007} argue that the probability of a hot Jupiter being trapped in an oblique Cassini state in the first place is small. Assuming the hot Jupiter does get trapped in an oblique Cassini state, \citet{Fabrycky2007} show that the evolution from Cassini state 2 to 1 (inclined spin axis to aligned spin axis) happens on a relatively fast timeline for hot Jupiters, and estimate that for HD 209458~b, by $\sim$35 Myr any spin axis inclination would be damped away. They therefore dismiss this for the field age of HD 209458, but for the young CoRoT-2, this may still be a possibility. Whether an inclined spin axis could explain the Western hotspot offset has not been as clearly shown in models, but some initial work suggests it's possible \citep{Rauscher2017, Ohno2019}. In both the inclined spin axis scenario and the sub-synchronous rotation scenario, CoRoT-2b is not in the aligned synchronous rotation configuration that is normally assumed (and not tidally locked). We therefore leave this as a potential interesting alternative to explore in future studies, but due to the larger body of work on sub-synchronous rotation causing a western hotspot offset and debated prospects of hot Jupiters being in oblique Cassini states, we keep sub-synchronous rotation as our preferred solution.

\subsubsection{Clouds}

The presence of clouds was also suggested as a mechanism responsible for the observed western hotspot offset. \citet{Dang2018} explain that clouds could obscure the emergent flux from the planet's eastern hemisphere and cause a western hotspot offset. However, it is thought that clouds preferentially form on the cooler hemisphere of the planet. Evidence for clouds on the cooler hemispheres of hot Jupiters has also been confirmed observationally \citep[e.g.,][]{Fu2025}. In normal circulation patterns, the western hemisphere is the cooler hemisphere and so clouds would form on the western side and not on the eastern side, as is required to explain CoRoT-2~b's observations. Therefore, either a western circulation pattern is again needed to create a cooler eastern hemisphere, or photochemical hazes (as opposed to standard condensate clouds) may form preferentially on the hotter hemisphere of the planet \citep{Kempton2017}. Although more complex treatment of hazes within GCMs finds that the expected distribution of hazes with longitude is more homogeneous than initially expected, and therefore preferentially forming hazes only on the hotter hemisphere of the planet may also be challenging \citep{Steinrueck2023}. 

Using the results from our post-eclipse retrievals on the CRIRES+ data, in combination with the pre-eclipse IGRINS retrievals presented in \citet{Shu2025}, we search for evidence of clouds in both datasets and aim to compare them. Neither high-resolution dataset returns strong constraints on cloud properties, which is unsurprising and a known limitation of the method due to limited wavelength coverage and the continuum information being removed during the data cleaning process in standard high-resolution analyses. However, even without strong constraints, neither high-resolution only retrieval show any clear degeneracies in the corner plots with the PT profile and both can rule out thick clouds above about $10^{-2}$ bars. 
For pre-eclipse phases (IGRINS), the maximum likelihood value for the cloud top location is between $10^{-0.06}$ and $10^{-0.47}$ (depending on the retrieval), while at post-eclipse phases (CRIRES+) the cloud top is located between $10^{-0.43}$ and $10^{-0.69}$. 

\citet{Shu2025} also run a joint retrieval between the high-resolution IGRINS data, in combination with $HST$ and $Spitzer$ data. By adding in the low-resolution spectra, they are able to derive a bounded constraint on the cloud-top pressure for the first time. 
The $HST$ data show a relatively featureless emission spectrum, which drives the bounded constraint on the cloud top pressure in the joint-retrieval, and finds a cloud top location between $10^{0.58}$ and $10^{-0.81}$ bars. Although the constraint derived when including the $HST$ and $Spitzer$ data is not phase-dependent, the consistency  between all the datasets suggests that clouds are not strongly affecting our spectra and that any clouds that are present do not strongly varying across the surface of the planet at phases probed by our observations. 

While our data points towards clouds not being the main culprit of the western hotspot offset, due to the heavy reliance on high-resolution near-infrared spectroscopy, we refrain from drawing strong conclusions from these particular observations. Low resolution observations at optical wavelength will be able to draw much stronger conclusions on the presence of clouds and hazes, and so we encourage future observations at shorter wavelengths in order to confirm or rule out clouds and hazes as the hotspot offset mechanism.

\section{Conclusions}
\label{s:conc}

Unlike most hot Jupiters which are measured to have either no hotspot offset or an eastern hotspot offset, CoRoT-2~b displays robust evidence for a western hotspot offset from a full $Spitzer$ phase curve \citep{Dang2018}. A western hotspot offset is unexpected, as it means that the hottest point on the planet is displaced in the $opposite$ direction of tidally locked rotation. \citet{Dang2018} proposed three possible scenarios that may cause the hotspot to be offset in the western direction: (1) magnetism, (2) sub-synchronous rotation, and (3) clouds. 

To attempt to confirm and determine the origin of this western hotspot offset, we present two new epochs of VLT/CRIRES+ spectra of CoRoT-2~b. These epochs were combined with a previous Gemini-S/IGRINS epoch to measure the planet's phase-resolved emission spectrum. 
We are able to significantly measure the signal of the planet at pre- and post-eclipse phases separately, and perform both a cross correlation analysis and a retrieval analysis to compare pre- and post-eclipse parameters. Our post-eclipse retrievals return consistent abundances with the pre-eclipse retrievals from \citet{Shu2025}, but we measure a hotter and more isothermal PT profile, independent of the parameterization of the PT profile used. This hotter and more isothermal PT profile is consistent with a western hotspot offset, although we refrain from saying that our study is independent confirmation of the western hotspot offset due to the large uncertainties on the PT profiles. 

We also used the data to understand the cause of the previously measured western hotspot offset. We do not find significant evidence for radial velocity offsets from the planetary rest frame as a function of phase, indicating a lack of evidence for strong winds in the direction of planetary rotation (i.e. lack of eastern super-rotational jet). Our pre- and post-eclipse retrievals also do not find any strong evidence for clouds, and both retrievals return similar (although poorly constrained) values for cloud top pressures around 0.1 to 1 bars. Finally, we measure a planetary rotational broadening of $2.24\substack{+0.81\\-0.77}$ km s$^{-1}$, which is smaller than our calculated velocity assuming synchronous rotation of the planet (4.37$\pm0.13$ km s$^{-1}$). This small rotational broadening is consistent with GCM models for expectations of sub-synchronous rotation and also points towards the existence of a western flow, which further narrows the planetary line profiles and serves to physically drive a western hotspot offset. 
Taken together, we find that our data are most consistent with the sub-synchronous rotation hypothesis due to the small planetary rotational velocity and lack of clear evidence pointing towards the other two hypotheses, although we stress the need for further data to confirm this. 
 
Overall our data demonstrate the power of high resolution spectroscopy for constraining atmospheric dynamics in hot Jupiters, and pave the way for a new era of atmospheric dynamics with ELTs.

\newpage

This work is based on observations collected at the European Southern Observatory (ESO) under program 111.24WF (PI:\,Kesseli) using the CRIRES$^{+}$ spectrographs on the ESO VLT.  
CRIRES$^{+}$ is an ESO upgrade project carried out by Thüringer Landessternwarte Tautenburg, Georg-August Universität Göttingen, and Uppsala University. The project is funded by the Federal Ministry of Education and Research (Germany) through Grants 05A11MG3, 05A14MG4, 05A17MG2 and the Knut and Alice Wallenberg Foundation.
This work also used The Immersion Grating Infrared Spectrometer (IGRINS) that was developed under a collaboration between the University of Texas at Austin and the Korea Astronomy and Space Science Institute (KASI) with the financial support of the US National Science Foundation under grants AST-1229522, AST-1702267 and AST-1908892, McDonald Observatory of the University of Texas at Austin, the Korean GMT Project of KASI, the Mt.\ Cuba Astronomical Foundation and Gemini Observatory. This work is based on observations obtained at the international Gemini Observatory, a program of NSF’s NOIRLab, which is managed by the Association of Universities for Research in Astronomy (AURA) under a cooperative agreement with the National Science Foundation on behalf of the Gemini Observatory partnership: the National Science Foundation (United States), National Research Council (Canada), Agencia Nacional de Investigación y Desarrollo (Chile), Ministerio de Ciencia, Tecnología e Innovación (Argentina), Ministério da Ciência, Tecnologia, Inovações e Comunicações (Brazil), and Korea Astronomy and Space Science Institute (Republic of Korea).

The preparation of this paper and of observation scheduling has made use of the NASA Exoplanet Archive, which is operated by the California Institute of Technology, under contract with the National Aeronautics and Space Administration under the Exoplanet Exploration Program.

This project has been carried out within the framework of the National Centre of Competence in Research PlanetS supported by the Swiss National Science Foundation under grant 51NF40\_205606. S.P.\ acknowledges the financial support of the SNSF.
L.D. ans Y.S. acknowledge support from the Natural Sciences and Engineering Research Council (NSERC) and the Trottier Family Foundation. V.P. would like to thank support from the UKRI STFC grant UKRI1171.

\bibliography{bib}{}
\bibliographystyle{aasjournal}

\appendix

\section{Retrieval Corner Plots}
\label{A:ret}

\begin{figure*}[h!]
\begin{center}
\includegraphics[width=\linewidth]{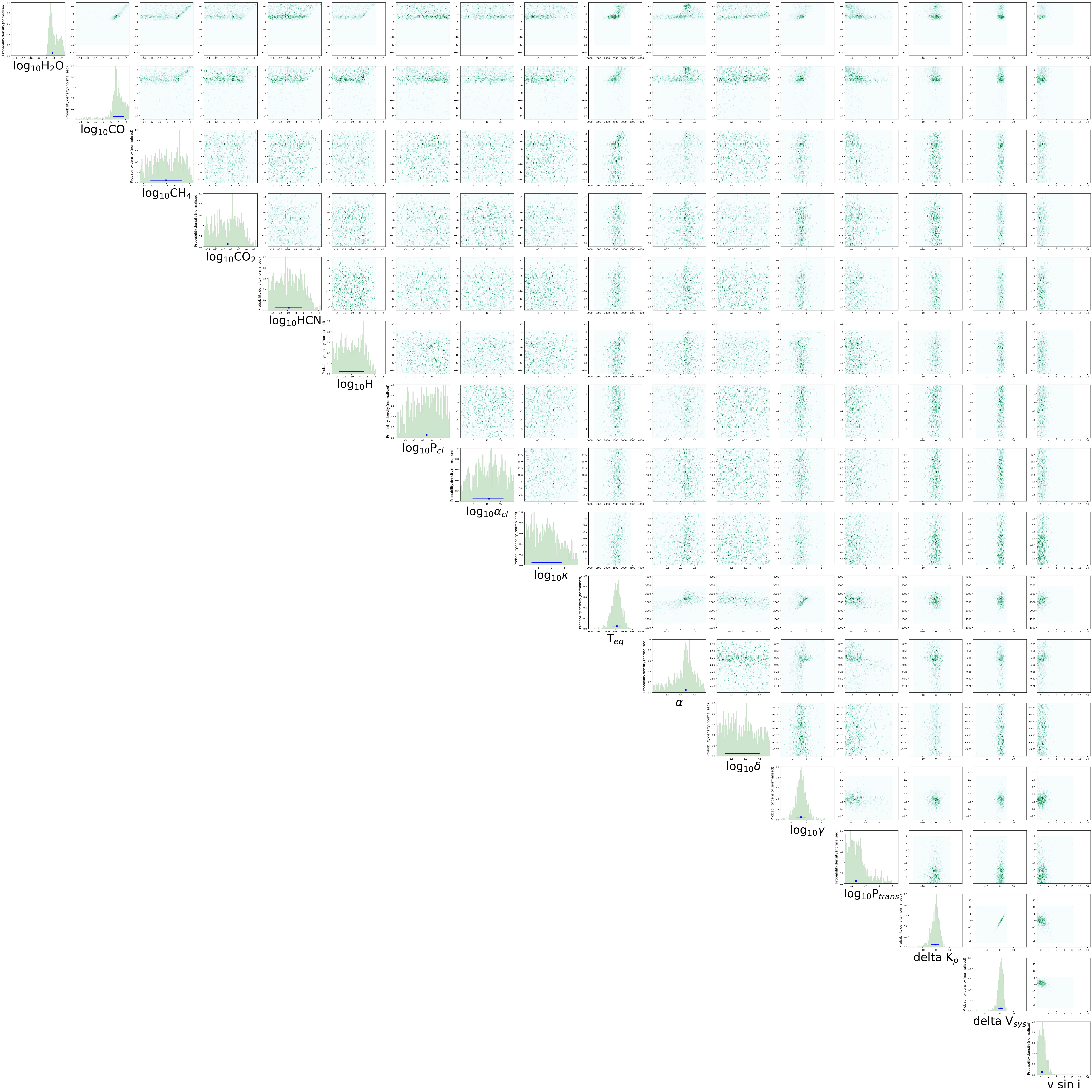}
\caption{\small 
Corner plot showing the posterior distribution of all inferred parameters from the nominal atmospheric retrieval, which included major C- and O-bearing species expected to exist in a CoRoT-2b-like planet. The model also includes a uniform gray cloud deck and the PT profile is parameterized by a modified Guillot PT profile \citep{Molliere2019, Guillot2011}. 
}
\label{f:corner_nom}
\end{center}
\end{figure*}

\begin{figure*}[h!]
\begin{center}
\includegraphics[width=\linewidth]{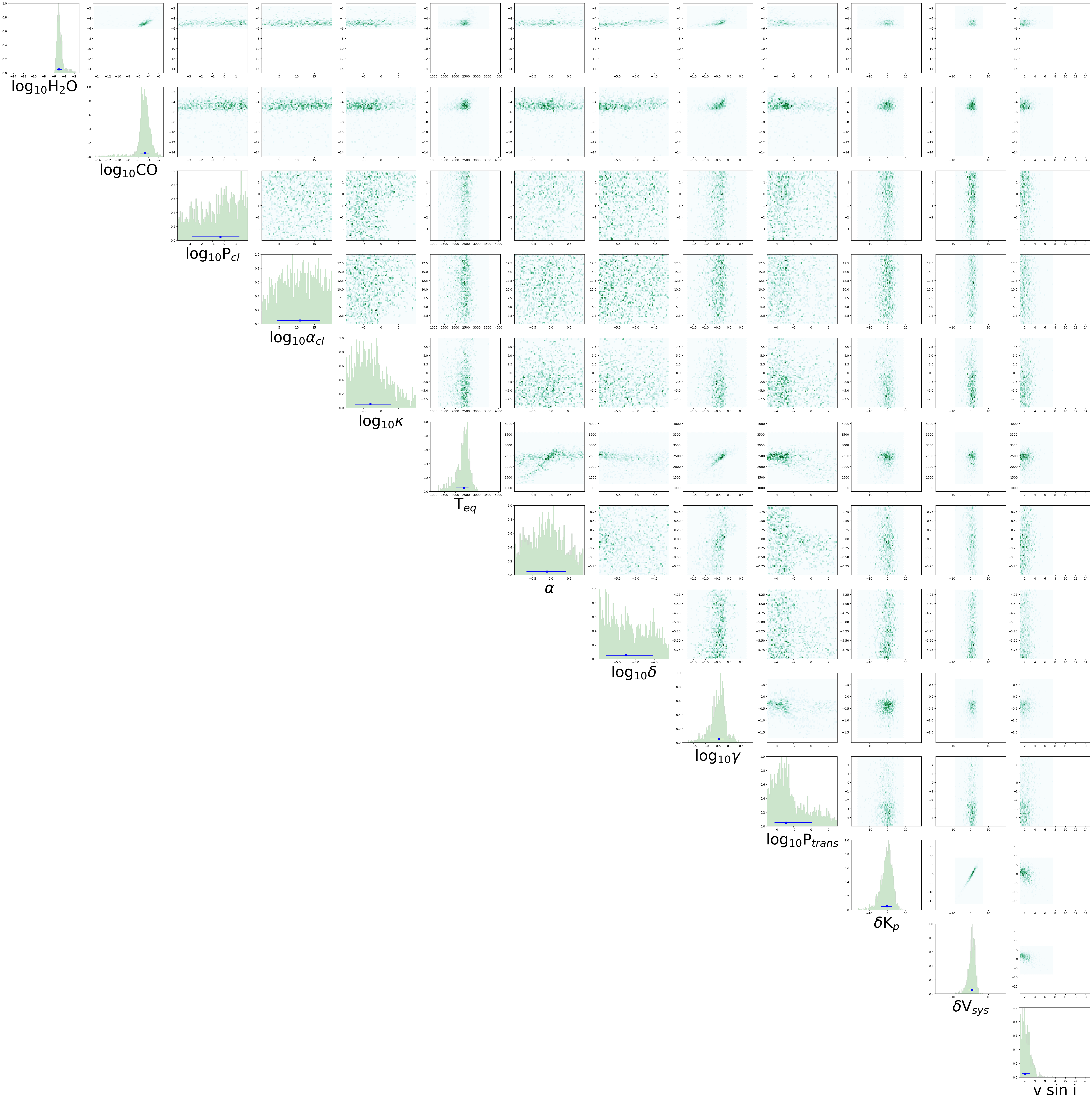}
\caption{\small 
Same as Figure \ref{f:corner_nom}, but only run with the two molecules detected in cross correlation (H$_2$O and CO). 
}
\label{f:corner_2species}
\end{center}
\end{figure*}

\begin{figure*}[h!]
\begin{center}
\includegraphics[width=\linewidth]{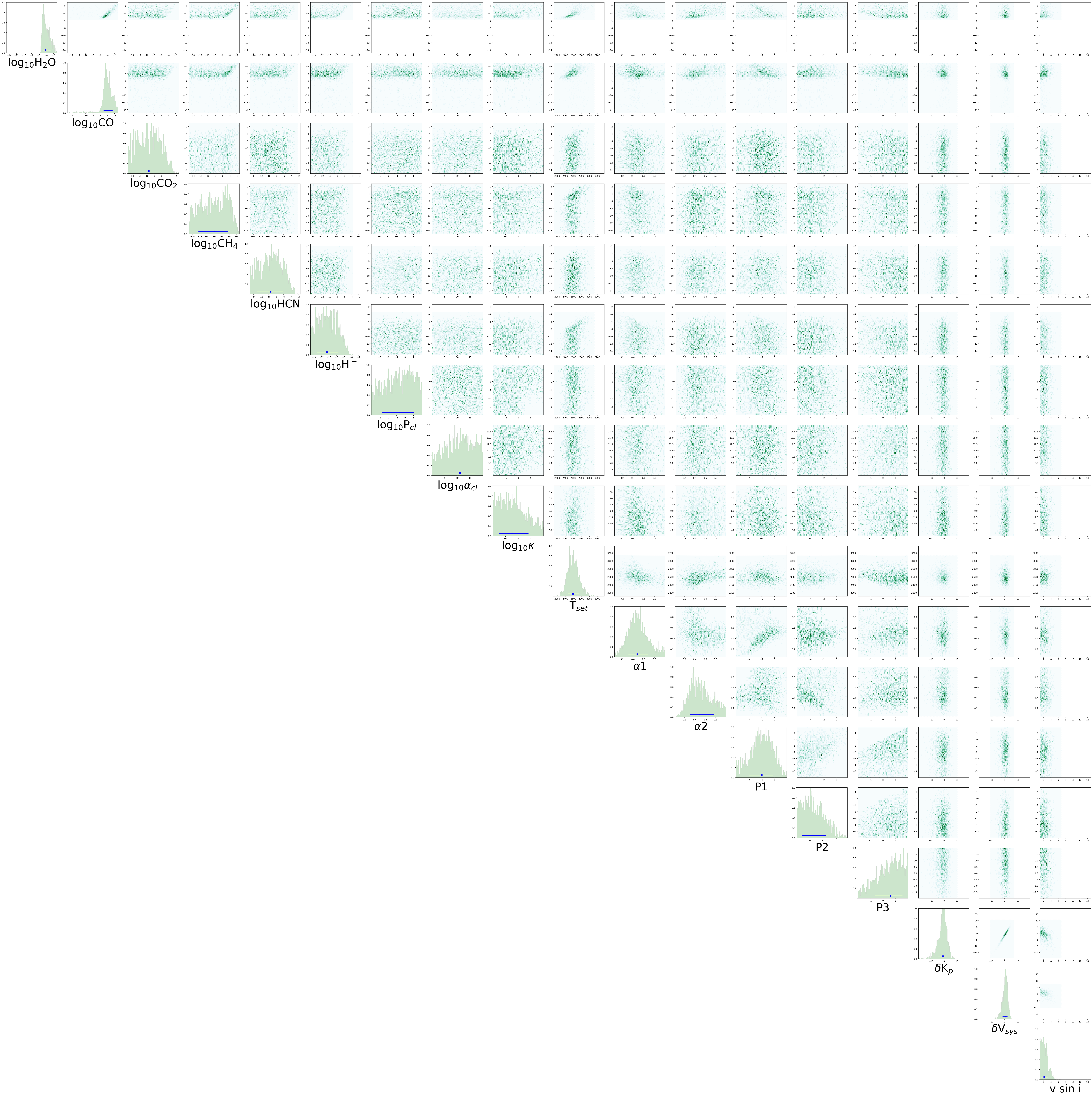}
\caption{\small 
Same as Figure \ref{f:corner_nom}, but the PT parameterization is instead given by \citet{Madhusudhan2009}.   
}
\label{f:corner_sidPT}
\end{center}
\end{figure*}

\section{Calculation for Tidal Locking}
\label{B: tidal_lock}

Using Equation 9 in \citet{Gladman1996}, we estimate the tidal locking timescale ($\tau$) for CoRoT-2~b. The equation can be written as: 

\begin{equation}
    \tau = \frac{\omega a^6 I Q}{3 G M_* \kappa_2 R_p^5}
\end{equation}

\noindent where $\omega$ is the initial spin rate (radians/s), $I$ is the moment of inertia and equal to $I\simeq0.4 M_p R_p^2$, $Q$ is the dissipation function, and $\kappa_2$ is the Love number. The other parameters ($a$, $M_*$, $M_p$, and $R_p$) are known values and taken from Table \ref{t:corot2}.  

We must estimate values for $\omega$, $\kappa_2$, and $Q$ since those are not known precisely for CoRoT-2b. We take $\omega$ as Jupiter's current spin rate ($1.76\times10^{-4}$ rad/s). Love numbers can range from 0 to 1.5, and have been measured for a few hot Jupiters \citep[see][for a review]{vandijk2025}.  Typical values for the measurements are $\sim$0.5, which we use in our calculation. The parameter with the largest uncertainty is $Q$, which is often taken to be 100 for unknown bodies, although estimates for Jupiter place it at $\sim10^7$ \citep{Wu2005}. Using all the stated values above and trying different values of Q between 100 and $10^7$ we find that at most the tidal locking timescale would be a few million years, which is consistent with estimates for similar systems like 51 Peg b \citet{Rasio1996}. With an age estimate of $100-300$ Myr, we conclude that CoRoT-2~b should have already synchronized.  


\end{document}